\documentclass[12pt,a4paper]{article}
\usepackage{setspace}
\usepackage[pdftex,colorlinks=true]{hyperref}
\usepackage{amsmath, appendix, bookmark, xcolor}
\usepackage[font=singlespacing]{caption}
\usepackage[top=1.4in, bottom=1.4in, left=1.2in, right=1.2in]{geometry}
\pdfmapfile{=gothict1.map}

\DeclareCaptionStyle{italic}[justification=centering]{labelfont={bf},font ={small},textfont={it},labelsep=colon}
\usepackage{graphicx,psfrag,epsf}
\usepackage{enumerate, algorithm, algorithmic}
\usepackage{natbib}
\usepackage{url} 
\usepackage{booktabs, subfig, bm, paralist, mathpazo, tikz, longtable, microtype, dsfont, rotating, multirow, ragged2e} 
\definecolor{darkblue}{rgb}{0,0,.6}
\hypersetup{citecolor=darkblue,linkcolor=darkblue,urlcolor=darkblue}

\newcommand{\blind}{0}

\newcommand{\X}{\mathcal{X}}
\newcommand{\Y}{\mathcal{Y}}

\graphicspath{{plots/}}

\newsavebox\CBox

\definecolor{a0}{rgb}{0.0, 0.5, 0.0}
\definecolor{bistre}{rgb}{0.24, 0.17, 0.12}
\definecolor{amethyst}{rgb}{0.6, 0.4, 0.8}
\definecolor{blue-violet}{rgb}{0.54, 0.17, 0.89}
\definecolor{Rcolor}{RGB}{150,160,190}
\definecolor{blush}{rgb}{0.87, 0.36, 0.51}
\definecolor{brightturquoise}{rgb}{0.03, 0.91, 0.87}
\definecolor{burntorange}{rgb}{0.8, 0.33, 0.0}
\date{\today}
\AtBeginDocument{\renewcommand{\harvardand}{and}}
\usepackage{orcidlink}

\newcommand{\appendixnumberline}[1]{Appendix\space}

\let\oldappendix\appendix
\makeatletter
\renewcommand{\appendix}{%
  \addtocontents{toc}{\let\protect\numberline\protect\appendixnumberline}%
  \renewcommand{\@seccntformat}[1]{Appendix~\csname the##1\endcsname\quad}%
  \oldappendix
}
\makeatother

\begin{document}

\def\spacingset#1{\renewcommand{\baselinestretch}
{#1}\small\normalsize} \spacingset{1}
\date{}

\if0\blind
{
  \title{\bf Stock Return Prediction based on a Functional Capital Asset Pricing Model}
  \author{
\normalsize    Ufuk Beyaztas \orcidlink{0000-0002-5208-4950} \\
\normalsize    Department of Statistics \\
\normalsize    Marmara University \\
\\
\normalsize Kaiying Ji \orcidlink{0000-0002-4332-3892} \\
\normalsize Discipline of Accounting, Governance and Regulation \\
\normalsize University of Sydney \\
  \\
\normalsize  Han Lin Shang \orcidlink{0000-0003-1769-6430}\footnote{Correspondence: Department of Actuarial Studies and Business Analytics, Macquarie University, NSW 2109, Australia; Telephone: +61(2) 9850 4689; Email: hanlin.shang@mq.edu.au}\\
\normalsize    Department of Actuarial Studies and Business Analytics \\
\normalsize    Macquarie University \\
    \\
\normalsize Eliza Wu \orcidlink{0000-0003-3937-8660} \\
\normalsize Discipline of Finance \\
\normalsize University of Sydney 
}
  \maketitle
} \fi

\if1\blind
{
  \title{\bf Stock Return Prediction based on a Functional Capital Asset Pricing Model}
  \author{}
  \maketitle
} \fi

\bigskip

\begin{abstract}
The capital asset pricing model (CAPM) is readily used to capture a linear relationship between the daily returns of an asset and a market index. We extend this model to an intraday high-frequency setting by proposing a functional CAPM estimation approach. The functional CAPM is a stylized example of a function-on-function linear regression with a bivariate functional regression coefficient. The two-dimensional regression coefficient measures the cross-covariance between cumulative intraday asset returns and market returns. We apply it to the Standard and Poor's 500 index and its constituent stocks to demonstrate its practicality. We investigate the functional CAPM's in-sample goodness-of-fit and out-of-sample prediction for an asset's cumulative intraday return. The findings suggest that the proposed functional CAPM methods have superior model goodness-of-fit and forecast accuracy compared to the traditional CAPM empirical estimation. In particular, the functional methods produce better model goodness-of-fit and prediction accuracy for stocks traditionally considered less price-efficient or more information-opaque.
\end{abstract}


\newpage
\spacingset{1.7}

\section{Introduction}\label{sec:1}

The capital asset pricing model (CAPM) \citep{S64, L65} has been the cornerstone of theoretical and empirical finance. It is a benchmark model in asset pricing research due to its simplicity and efficiency in measuring the cost of equity. CAPM assumes that securities markets are highly competitive and efficient, so relevant information about companies is quickly and universally distributed and absorbed. It also assumes that these markets are dominated by rational, risk-averse investors who seek to maximize returns on their investments; that is, they demand higher returns for greater risks. Therefore, asset prices take into account all available information in an efficient market without arbitrage opportunities, and the expected return of a stock is influenced by its correlation with the market index, which is measured by beta.

Since the introduction of CAPM, much attention has been placed on improving its model specifications \citep[see, e.g.,][]{CV82, JM02, BCG07}. For example, considerable evidence suggests that elements in the CAPM model, that is, market risk premium, risk-free interest rates, and betas, are time-varying \citep[see, e.g.,][]{BEW88, FF89, DBO89, JW96, G98, Wang03, AC03, ZP17}, and conditional modeling that explicitly allows for temporal variation in the factor loading results in statistically significant and economically meaningful improvements \citep[see, e.g.,][]{BK89, BBP88, w92, BNS95, E96, F02, ABDL03, BCT09, BET17, ZP17}. \cite{BET17} argue that generalized autoregressive conditional heteroskedasticity-based time-varying conditional betas help explain cross-sectional variation in expected stock returns.

Building on the conditional CAPM, \cite{BLT16} use high-frequency-based estimates and propose an asset pricing framework that uses a continuous beta to reflect smooth intraday co-movements in the market and two rough betas associated with intraday price discontinuities, or jumps, during the active part of the trading day and overnight close-to-open return, respectively. Furthermore, \cite{BPQ22} propose ``granular betas" that provide a much more refined look at the inherent dependencies between an asset and a given set of factors. 

The literature on time-varying beta presents many models that consider the time at which closing prices are observed. Still, these models largely overlook changes in the intraday price curves, that is, how a financial instrument shifts from time to time. High-frequency financial data, that is, observations on financial instruments taken at a granular scale, such as time-stamped transaction-by-transaction or tick-by-tick data, may better address this deficiency \citep[see, e.g.,][]{BHR14, BCM+18, CMW22}. With the advent of new exchanges, every online transaction is recorded and compiled into databases, such as Island electronic communication networks or data for individual bids to buy and sell. 

Recent advances in computer storage and data collection have enabled researchers in finance to record and analyze high-frequency data to understand market microstructure related to price discovery and market efficiency \citep[see, e.g.,][]{CLM97, Engle00, Andersen00, ABDL01, ABDL03, GO97, Ghysels00, Wood00, DGMOP01, GJ01, Lyons01, AG02, DGMOP01, Tsay10, P13, GKG14, BPQ22}. This stream of research demonstrates that high-frequency-based sampling allows more accurate factor representations and improved asset pricing predictions compared to conventional lower-frequency estimates, resulting in more efficient ex-post mean-variance portfolios \citep{BZ03, CLRW16, HPS20}. \citet[][p.13]{ABDW04} provide further evidence suggesting that high-frequency beta measures are capable of "more clearly highlighting the dynamic evolution of individual security betas" compared to the results obtained from lower frequency daily data. \cite{AHHH12} introduces a modified functional CAPM and sequential monitoring procedures and suggests that the functional data-analytic approach performs better in detecting a structural break and capturing time variability in the betas.

High-frequency intraday financial data are examples of dense functional data in statistics, represented in graphical form as curves \citep[see, e.g.,][]{AST+23}. As an integral part of functional data analysis \citep{RS05, FV06} and time series analysis \citep{KR17, PS21}, functional time series consist of random functions observed at a time interval. Functional time series can be grouped into two categories, regardless of whether the continuum is also a time variable. On the one hand, functional time series can arise from measurements obtained by separating an almost continuous time record into consecutive intervals, for example, days, weeks, or years \citep[see, e.g.,][]{HK12}. We refer to such data structures as \textit{sliced} functional time series, examples of which include the intraday price or volatility curves of a financial stock \citep[see, e.g.,][]{Shang17b, SYK19, ATT21}. On the other hand, the functional variable can be other continuous variables, such as maturity-specific yield curves \citep[see, e.g.,][]{HSH12} or near-infrared spectroscopy wavelength \citep[see, e.g.,][]{SCS22}. In either case, the object of interest is a discrete-time series of functions with a continuum \citep[see, e.g.,][]{WJS08, KZ12, KRS17}. In Section~\ref{sec:2}, we describe our motivating data sets -- 5-minute cumulative intraday returns (CIDRs) of the Standard and Poor's 500 (S\&P 500) indexes and their constituents. These data sets belong to the first type of functional time series.

If prices are modeled as a univariate time series of discrete observations, the underlying process that generates these observations cannot be fully discovered. The advantages of functional time series include:
\begin{inparaenum}
\item[1)] Thanks to continuity, we can study the temporal correlation between two intraday functional objects.
\item[2)] The beta function estimate is a two-dimensional image capturing (cross-) correlation between an asset and its stock index to study the cross-sectional dependence between two random points on a functional object.
\item[3)] We handle missing values via interpolation or smoothing techniques.
\item[4)] By converting a univariate time series to a time series of functions, we implicitly overcome the ``curse of dimensionality" \citep{Donoho00}, where nonparametric and semiparametric techniques can be implemented \citep[see, e.g.,][]{FV06, AV06}.
\end{inparaenum}

We propose an extension of a CAPM tailored for high-frequency financial data, termed a functional CAPM. The functional CAPM has recently been considered in a working paper of \cite{Pedersen22}. Still, we present a novel way of estimating the regression coefficient and investigating the difference between the CAPM and its functional version from an aspect of firm characteristics. It is designed to explain how much variability (i.e., information) in the market cumulative intraday return can explain the variability in an asset. Our functional CAPM can be cast as a function-on-function linear regression, where our objective is to estimate the bivariate functional regression coefficient. The bivariate functional regression coefficient measures a linear relationship between a functional response (i.e., CIDRs of an asset) and a functional predictor (i.e., CIDRs of a market index). Through the functional CAPM, we can predict the in-sample conditional expectation of the CIDRs of an asset with the estimated bivariate regression coefficient function and evaluate the model's goodness of fit via a functional $R^2$. The findings in this study suggest that the proposed functional CAPM methods present superior performance in model goodness-of-fit for those less price-efficient stocks and better prediction accuracy for information opaque stocks.

Our article is structured as follows. Section~\ref{sec:2} describes a financial stock market index and its respective constituent stocks. Section~\ref{sec:fcapm} introduces the functional CAPM to estimate the bivariate regression coefficient function, which captures the linear relationship between the S\&P 500 index and its constituent stock. We apply the functional CAPM to model CIDRs by selecting assets representing various asset classes. In Section~\ref{sec:4}, we display the estimated regression coefficient functions obtained from the functional principal component regression (FPCR), functional partial least squares regression (FPLSR), and penalized function-on-function regression (PFLM). Intraday~$R^2$, root mean squared error (RMSE), and root mean squared prediction error (RMSPE) are presented to summarize the goodness-of-fit and out-of-sample prediction for the intraday financial data. The integrated values of these errors can be used to evaluate and compare the overall goodness-of-fit among different constituent stocks in Sections~\ref{sec:5.1} and~\ref{sec:5.2}. In Sections~\ref{sec:5.3} and~\ref{sec:5.4}, we relate the differences in forecast accuracy between the classical CAPM and functional CAPM with firm characteristics and corporate governance, respectively. Section~\ref{sec:6} concludes and offers some ideas on how the methodology presented can be further extended. Details of the FPCR, FPLSR, and PFLM techniques are presented in Appendixes~\ref{sec:app_a}-\ref{sec:app_c}.

\section{Empirical data analysis}\label{sec:2}

\subsection{Data and sample selection}\label{sec:2.1}

The S\&P500 (SPX) is a financial stock market index that tracks the performance of around 500 large companies listed on stock exchanges in the United States. The intraday tick history for the S\&P500 (SPX) index and its constituent stocks were obtained from the Refinitiv Datascope (\url{https://select.datascope.refinitiv.com/DataScope/}). We consider daily cross-sectional returns from January 4, 2021, to December 31, 2021. In 2021, there were 252 trading days. Following early work by \cite{BZ03}, we downloaded transaction prices for SPX and each constituent stock at each 5-min interval from 09:30 to 16:00. For each stock, the sampling process yielded 78 data points per day and approximately 20,000 data points in the sampling period for analysis. This study utilized 10 million intraday data points for demonstration. The sample size could be expanded without computing power constraints.

We obtained data on the financial balance sheet and the firm Global Industry Classification Standard (GICS) sector from Compustat. Daily data on returns, prices, market capitalizations, and volumes were obtained from the Center for Research in Security Prices (CRSP). Institutional holdings were downloaded from 13F filings. Analyst following and earnings forecast accuracy data were downloaded from the $I/B/E/S$ summary file, and board structure information was obtained from the BoardEx database (\url{https://wrds-www.wharton.upenn.edu/pages/about/data-vendors/boardex/}). Winsorization was not performed since dealing with extreme observations is a part of the functional modeling.
\begin{table}[!htb]
\centering
\tabcolsep 0.16in
\caption{Summary statistics of the S\&P 500 index and firm characteristics.}\label{tab:0}
\begin{small}
\begin{tabular}{@{}lrrrrrr@{}}
\toprule
Variables & Min & 1$^{\text{st}}$ Quartile & Median & Mean & 3$^{\text{rd}}$ Quartile & Max  \\
\midrule
SPX intraday price & 3667 & 4074 & 4298 & 4274 & 4488 & 4808  \\
\midrule
$\ln \text{MC}_{i,t}$ & 7.889 & 9.843 & 10.434 & 10.555 & 11.139 & 14.659 \\ 
$\ln \text{P}_{i,t}$ & 1.778 & 4.164 & 4.781 & 4.799 & 5.436 & 8.112 \\ 
$\text{LEV}_{i,t}$ & 0.102 & 0.505 & 0.653 & 0.640 & 0.774 & 0.990 \\ 
$\text{VOL}_{i,t}$ (million) & 2.191 & 20.075 & 40.540 & 91.834 & 89.521 & 1900.558 \\ 
$\text{ILLIQ}_{i,t}$ ($10^{-9}$) & 0.001 & 0.031 & 0.063 & 0.084 & 0.108 & 1.872 \\ 
$\text{BidAskSpread}_{i,t}$ & -8.452 & -7.991 & -7.806 & -7.758 & -7.622 & -4.361 \\ 
\midrule
$\text{Coverage}_{i,t}$ & 0.2624 & 2.5871 & 2.8502 & 2.8059 & 3.1061 & 3.9269\\
$\text{Accuracy}_{i,t}$ & 0.0000 & 0.0314 & 0.0957 & 0.3114 & 0.2962 & 6.0809 \\
$\text{InstoHold}_{i,t}$ & 0.002 & 0.728 & 0.829 & 0.846 & 0.907 & 3.763 \\ 
$\text{Independent}_{i,t}$ & 0.5556 & 0.8333 & 0.9000 & 0.8723 & 0.9167  & 1.0000 \\
*$\text{Duality}_{i,t}$ & 0.0000 & 1.0000 &  1.0000 & 0.8187 & 1.0000  & 1.0000  \\
\bottomrule
 \end{tabular}
\end{small}
\begin{justify}
\footnotesize{* Binary variable, there are 411 ones and 91 zeros.
}
\end{justify}
\end{table}

Table~\ref{tab:0} presents summary statistics for the S\&P 500 mid-point 5-minute intraday prices observed over 252 days. We also show the summary statistics of the firm characteristics; detailed variable definitions are presented in Section~\ref{sec:5.2}.

\subsection{Cumulative intraday returns of S\&P 500}\label{sec:2.2}

We considered five-minute resolution data, $78$ data points, covering the period from 9:30 to 16:00 Eastern Standard Time. For each asset, intraday 5-minute close price data, $P_t(u_i)$, are available on each trading day, which we used to construct a sequence of CIDRs \citep[see also][]{RWZ20}
\begin{equation}
\Y_t^j(u_i) = 100 \times [\ln P_t^j(u_i) - \ln P_t^j(u_1)], \qquad i=2,3,\dots,78, \label{eq:1}
\end{equation}
where $j$ denotes an asset in the S\&P 500 index, $i$ denotes the $i$\textsuperscript{th} intraday period, $t$ denotes a given trading day and $\ln(\cdot)$ denotes a natural logarithm. From~\eqref{eq:1}, we take the inverse transformation to obtain the 5-minute intraday price
\begin{equation*}
P_{t}^{j}(u_i) = \exp^{\frac{\Y^{j}_t(u_i)}{100}}\times P_{t}^{j}(u_1),
\end{equation*}
where $P^{j}_{t}(u_1)$ denotes the beginning close price on day $t$. 

In a given day~$t$, we observe CIDRs of the S\&P 500 index and its constituent stocks over the intraday period. Let $\X_{t}(u)$ be the functional predictor, consisting of the CIDR of a market index. Let $\Y^j_t(v)$ be the functional response, consisting of the CIDR of an index's constituent stock. Let $\bm{\Y}^j(v)=[\Y_1^j(v),\Y_2^j(v),\dots,\Y_n^j(v)]^{\top}$ and $\bm{\X}(u)$ = $[\X_1(u),\X_2(u),\dots,\X_n(u)]^{\top}$ be two functional time series of response and predictor, respectively. Let $\bm{\Y}^{j,c}(v) = \bm{\Y}^j(v) - \bm{R}^f$ and $\bm{\X}^{c}(u) = \bm{\X}(u) - \bm{R}^f$ denote excess intraday returns of an asset and a stock, where $\bm{R}^f=[R^{f}_{1}, R^{f}_{2},\dots, R^{f}_{n}]^{\top}$. $R_t^f$ can be computed by dividing the daily treasury par yield curve rate at one-year maturity by $(251 \times 78)$ intraday trading time over 251 trading days. 

\section{Functional Capital Asset Pricing Model (CAPM)}\label{sec:fcapm}

Before presenting the proposed functional CAPM, we provide an intuitive understanding of how the curves are obtained from discrete data.

\subsection{From discrete data points to curves}\label{sec:B-spline}

With high-frequency financial data, the data points are observed discretely. Our intraday financial data are observed densely at an equally-spaced grid, a five-minute time interval. A basis function expansion can convert the discrete data points into a continuous function. Because of this, an advantage of our functional data-analytic approach is that it can address non-synchronicities, that is, time intervals of irregular lengths, in asset returns when measured at high frequencies \citep{Dimson79, LN06, GHK+14, BCF+16}.

Among all possible basis functions, those widely used are polynomial basis functions (which are constructed from the monomials $\phi_k(u)=u^{k-1}$), Bernstein polynomial basis functions (which are constructed from 1, $1-u$, $u$, $(1-u)^2$, $2u(1-u)$, $u^2$, \dots), Fourier basis functions (which are constructed from 1, $\sin(\omega u)$, $\cos(\omega u)$, $\sin(2\omega u)$, $\cos(2\omega u)$, \dots), radial basis functions, wavelet basis functions, spline basis functions, and orthogonal basis functions. Our functional predictor and response are approximated by 20 $B$-spline basis functions:
\begin{align}
\Y_t^{j,c}(v) &= \sum_{k=1}^{20} \widehat{z}^j_{t,k} \widehat{\pi}^j_k(v) = \bm{\widehat{Z}}_t^j \bm{\widehat{\Pi}}^j(v), \label{eq:B_spline_1}\\
\X_t^{c}(u) &= \sum_{m=1}^{20} \widehat{a}_m \widehat{\gamma}_m(u) = \bm{\widehat{A}}_t \bm{\widehat{\Gamma}}(u), \label{eq:B_spline_2}
\end{align}
where $\bm{\widehat{\Pi}}^j(v)$ and $\bm{\widehat{\Gamma}}(u)$ are the bases, and $\bm{\widehat{Z}}$ and $\bm{\widehat{A}}$ are the corresponding coefficient matrices. 

This \textit{pre-smoothing} step allows us to mitigate the curse of dimensionality by choosing to work in a square-integrable function space. From these discrete data points, a continuous curve can be constructed from $[\X_t(u_2),\dots, \X_t(u_{78})]$ representing six and half hours of trading at the New York Stock Exchange on a trading day.

\begin{figure}[!htb]
\centering
\includegraphics[width=8.4cm]{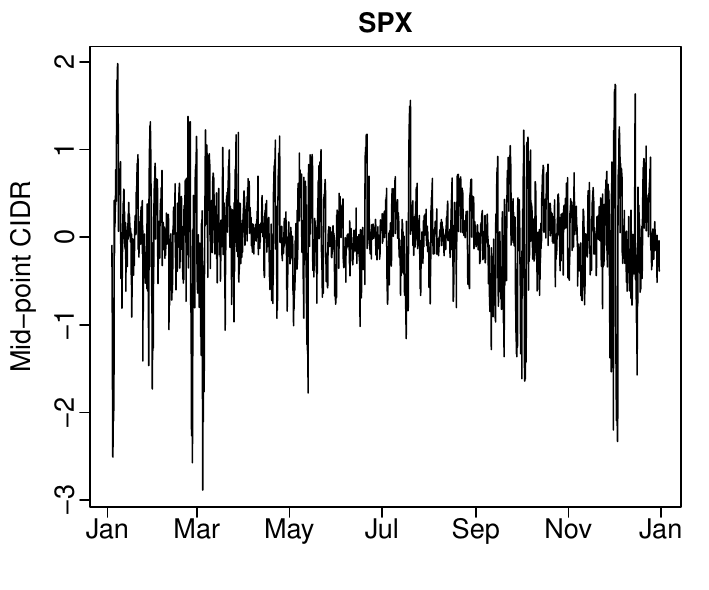}
\quad
\includegraphics[width=8.4cm]{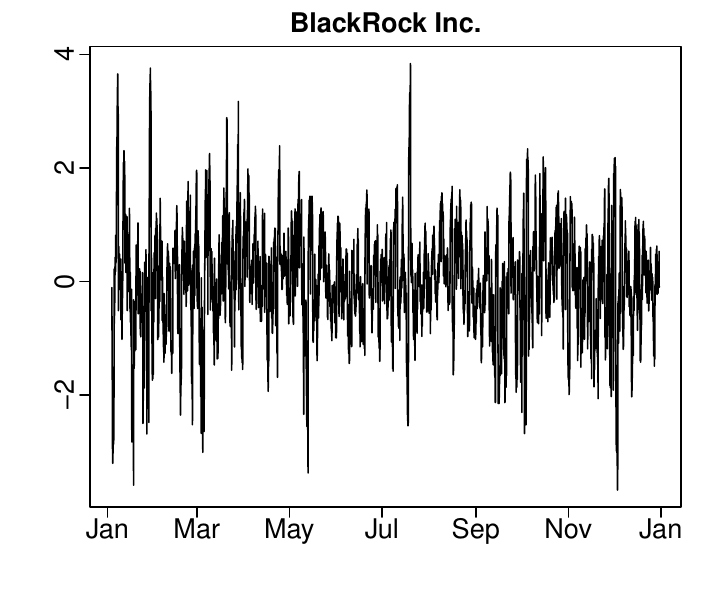}
\\
\includegraphics[width=8.4cm]{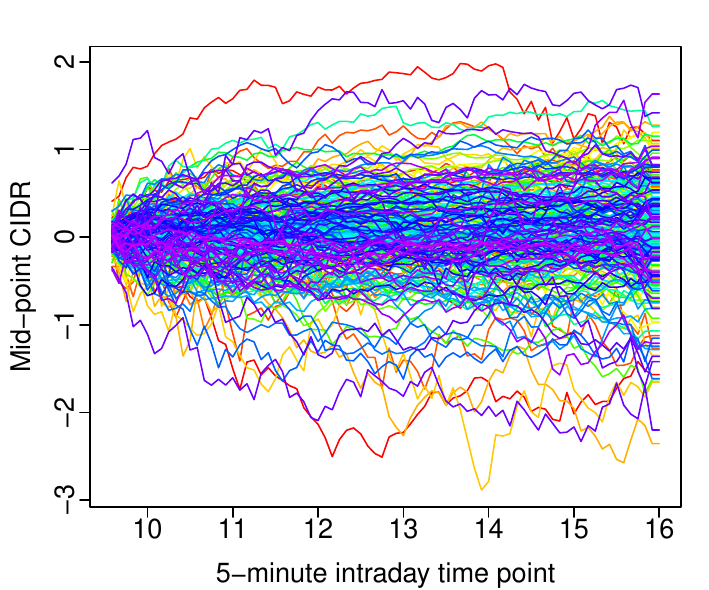}
\quad
\includegraphics[width=8.4cm]{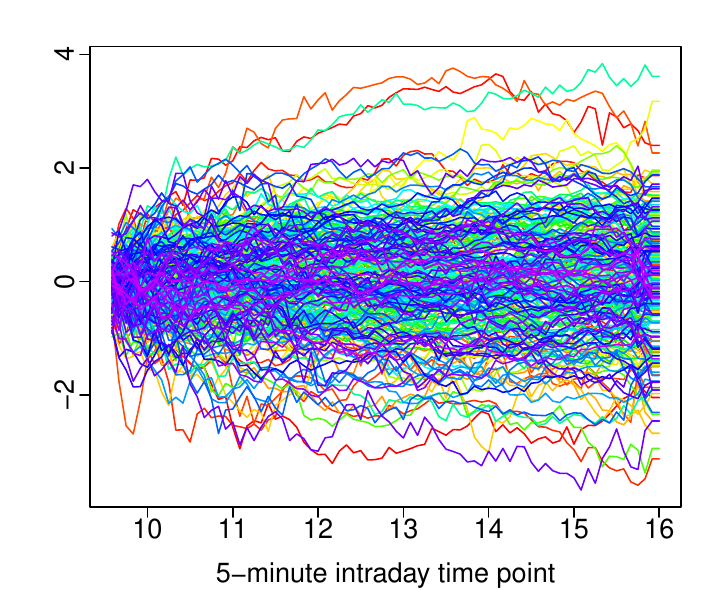}
\caption{Univariate and functional time series plots of the CIDRs of the SPX index and BlackRock Inc.}\label{fig:1}
\end{figure}

A univariate time series of $19,327$ discrete returns was converted into $n=251$ days of CIDR curves. In Figure~\ref{fig:1}, we present CIDRs for the S\&P 500 and BlackRock Inc. Using a functional KPSS test of \cite{HKR14}, both CIDR series are stationary with $p$-values of 0.669 and 0.546, respectively.

\subsection{Functional CAPM}\label{sec:3}

The classic CAPM can be expressed as:
\begin{equation*}
R_i = R_f + \beta_i (R_m - R_f) + \vartheta_i,
\end{equation*}
where $R_i$ and $R_m$ denote daily asset and market returns, $R_f$ denotes a risk-free rate of interest, $\beta_i$ is a real-valued slope parameter associated with the asset, and $\vartheta_i$ denotes an error term with a mean of zero and finite variance.

The functional CAPM captures a linear relationship between a centered functional predictor and a centered functional response through an unknown bivariate regression coefficient function, known as beta surface. The functional CAPM can be expressed as
\begin{equation}
\Y_t^{j}(v) = R_t^f + \int_{\mathcal{I}} \beta^j(u, v) \big[\X_t(u) - R_t^f\big]du + \varepsilon_t^j(v),\label{eq:CAPM}
\end{equation}
where $R_t^f$ denotes the intraday risk-free rate of interest, $\X_t^{c}(u) = [\X_t(u) - R_t^f]$ denotes the intraday market risk premium, $\beta^j(u,v)$ is the bivariate regression coefficient function associated with the $j$\textsuperscript{th} constituent stock, $\varepsilon_t^j(v)$ denotes an independent and identically distributed (i.i.d.) random error function, and $u, v\in \mathcal{I}$ denote a function support range of $\mathcal{I}$ (i.e., intraday trading period between 9:30 and 16:00 Eastern Standard Time). 

In~\eqref{eq:CAPM},  we model the relationship between the functional response $\Y(v)$ and the functional predictor $\X(u)$ via a bivariate regression coefficient $\beta(u,v)$. To prevent any reliance on future information (look-ahead bias), we adjust the limits of integration such that, for a given $v$ in the interval $\mathcal{I}$,  the integral on the right-hand side of~\eqref{eq:CAPM} only includes values of $\X(u)$ for $u \leq v$. This ensures that when predicting $\Y(v)$, only information from $\X$ up to the current time $v$ is used, thus preserving the causal nature of the model.

The functional CAPM is a special case of the concurrent function-on-function linear regression \citep[see, e.g.,][]{RD91}. The direct estimation of the regression coefficient in the functional CAPM is an ill-posed problem due to the singularity and curse of dimensionality. Since the functional predictor and response belong to infinite-dimensional function space, we consider projecting the functional predictor and response onto orthonormal and $B$-spline bases.

\subsection{Estimation of regression coefficient function}

Within the framework of functional CAPM in~\eqref{eq:CAPM}, it is crucial to accurately estimate the bivariate regression coefficient function $\beta^j(u,v)$ from a finite sample. Toward this end, we explore three distinct methodologies: FPCR, FPLSR, and PFLM. The PFLM relies on general basis expansion techniques such as $B$-spline basis functions. The FPCR and FPLSR adopt a data-driven dimension reduction paradigm, and they entail projecting infinite-dimensional curves onto finite-dimensional spaces of orthonormal bases. In contrast, PFLM may necessitate a larger number of basis functions to approximate the functional regression coefficient, potentially leading to model overfitting and reduced prediction accuracy.

In FPCR, the components used for approximating the bivariate regression coefficient function $\beta^j(u,v)$ are derived from the covariance among functional predictors alone. A few leading principal components comprise most of the variance between the functional predictors. These latent components may not necessarily be important to prediction accuracy \citep[see, for example][]{Delaigle2012}. FPLSR addresses this issue by leveraging both response and predictors while extracting latent components, thereby capturing more information with comparably fewer terms. Additionally, studies have demonstrated that FPLSR offers more accurate parameter function estimation compared to FPCR, albeit with greater dimension reduction \citep[see, e.g.,][]{Aguilera2010, BS2022, SBAS2022}.

In both FPCR and FPLSR, the data-driven orthogonal bases may lack smoothness to the functional parameter. However, this can lead to significant under-smoothing if the functional parameter exhibits considerably more smoothness than the higher-order FPLSR and FPCR scores \citep[see, e.g.,][]{ivanescu2015, BHS2024}. Consequently, including one or two additional latent components may alter the shape and interpretation of the functional parameter \citep[see also][]{Crainiceanu2009}. Contrarily, in PFLM, the penalty term applied during the estimation phase imposes a specific level of smoothness on the parameter estimate, thereby preventing overfitting.

The comprehensive details regarding the methodologies of FPCR, FPLSR, and PFLM for estimating the bivariate regression coefficient function $\beta^j(u,v)$ in~\eqref{eq:CAPM} have been deferred to Appendixes~\ref{sec:app_a}-\ref{sec:app_c}.

\section{Illustration of the functional CAPM}\label{sec:4}

\begin{figure}[!htb]
\centering
\includegraphics[width=5.45cm]{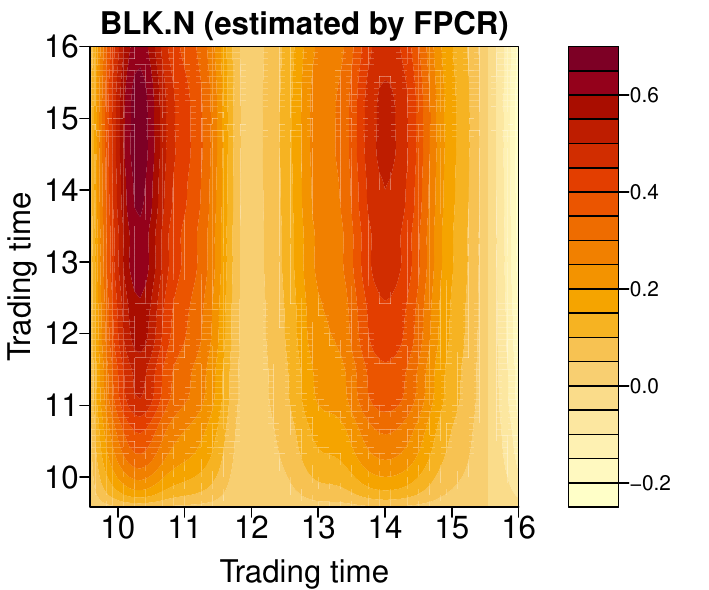}
\quad
\includegraphics[width=5.45cm]{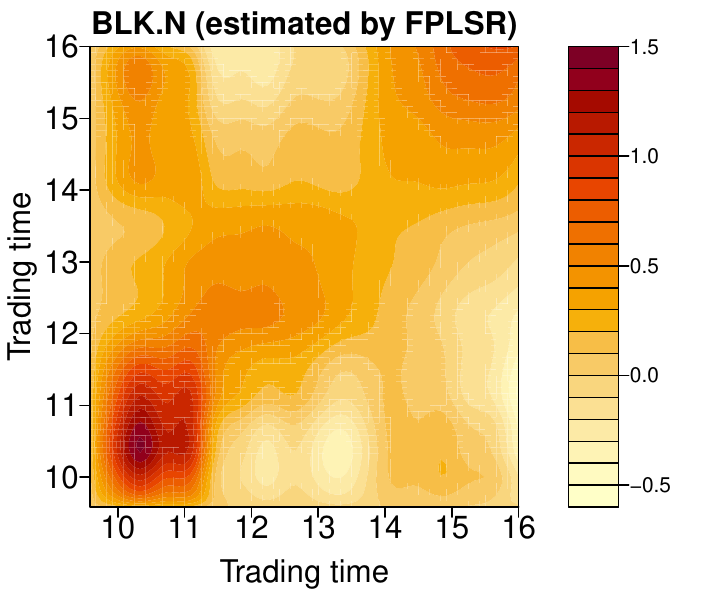} 
\quad
\includegraphics[width=5.45cm]{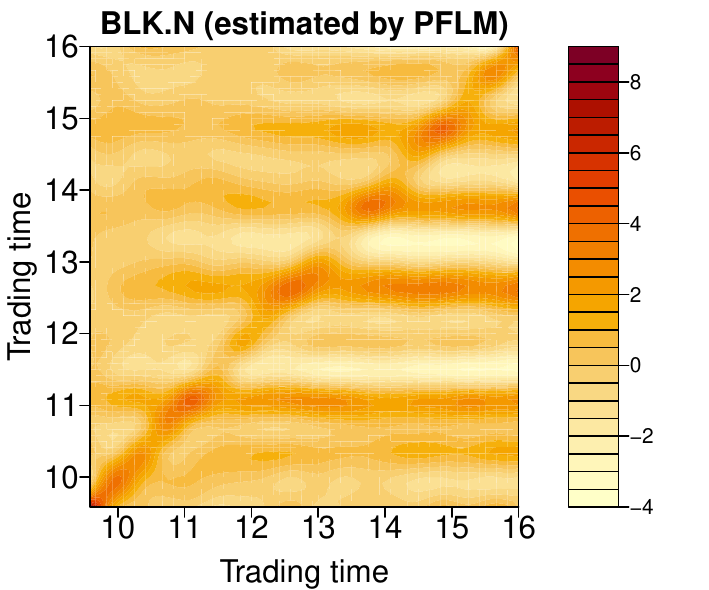} 
\caption{Plots of the estimated bivariate regression coefficient function $\widehat{\beta}^j(u,v)$ when the CIDR curves of a market index are the realizations of the functional predictor and the CIDRs of an asset are the functional response realizations. The bivariate regression coefficient function is estimated via the FPCR, FPLSR, or PFLM on the left, bottom, and right panels, respectively.}  \label{fig:2}
\end{figure}

We apply the FPCR and FPLSR to estimate the bivariate regression coefficient function in the functional CAPM. We consider the CIDRs of the S\&P 500 index as the functional predictor and of BlackRock Inc. stock as the functional response for demonstration. In Figure~\ref{fig:2}, we display the estimated regression coefficient functions obtained from the two function-on-function regression models. The regression coefficient function, estimated by the FPCR, shows intense activity between 10:00 and 11:30 and between 13:30 and 14:30. In contrast, the regression coefficient function estimated by the FPLSR demonstrates intense activity between 10:00 and 11:30 and between 15:00 and 16:00.

The difference between the estimated beta surfaces is because of the characteristics of the basis functions. In the FPCR, the basis functions obtained from the functional predictors are orthonormal, and there are no off-diagonal elements. Thus, its estimated beta surface is comparably smooth. In contrast, the FPLSR basis functions obtained from the functional predictors and responses are not orthogonal. The inverse square root of the inner product matrices plays an essential role in computing the regression coefficient surface of the FPLSR. In the FPLSR basis functions, the off-diagonal elements of the inner product matrices capture the linear dependence between the functional predictor and response at different intraday periods. As a result, the FPLSR can show more local features than the FPCR.

\section{Estimation accuracy of the response}\label{sec:5}

Using BlackRock Inc.'s CIDRs, we evaluate and compare the model performance to a traditional CAPM estimation, FPCR, and FPLSR. We firstly measure the in-sample goodness-of-fit and estimation accuracy of FPCR and FPLSR by computing $R^2$ and root-mean-square error (RMSE) and then compute root-mean-square percentage error (RMSPE) to measure one-step-ahead out-of-sample prediction accuracy. Further, we investigate whether firm characteristics, such as industry sector, firm size, leverage, liquidity, and valuation uncertainty impact these performance measures.

\subsection{In-sample goodness-of-fit}\label{sec:5.1}

While the estimated regression coefficient functions differ, we compute an intraday version of $R^2$ and RMSE as two in-sample goodness-of-fit criteria. The intraday $R^2$ and RMSE criteria extend from conventional linear models, defined as
\begin{align*}
R^2(v) &= 1 - \frac{\sum^n_{t=1}[\Y_t^j(v) - \widehat{\Y}_t^j(v)]^2}{\sum_{t=1}^n[\Y_t^j(v) - \overline{\Y}^j(v)]^2}, \qquad v\in \mathcal{I}, \\
\text{RMSE}(v) &= \sqrt{\frac{1}{n}\sum^n_{t=1}[\Y_t^j(v) - \widehat{\Y}_t^j(v)]^2} \\
&= \sqrt{\frac{1}{n}[1-R^2(v)]\sum^n_{t=1}[\Y_t^j(v) - \overline{\Y}^j(v)]^2},
\end{align*}
where $\widehat{\Y}_t^j(v)$ represents the fitted values obtained from the functional CAPM, using the estimated regression coefficient function. 

\begin{figure}[!htb]
\centering
\includegraphics[width=8.46cm]{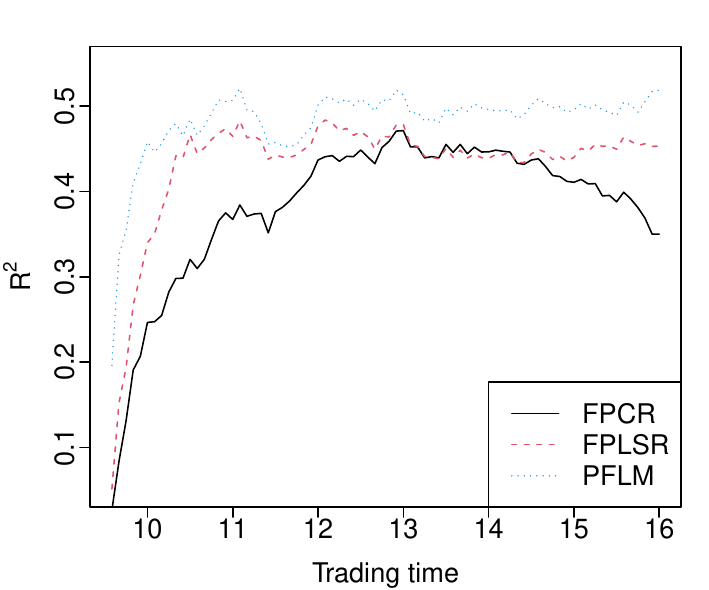}
\quad
\includegraphics[width=8.46cm]{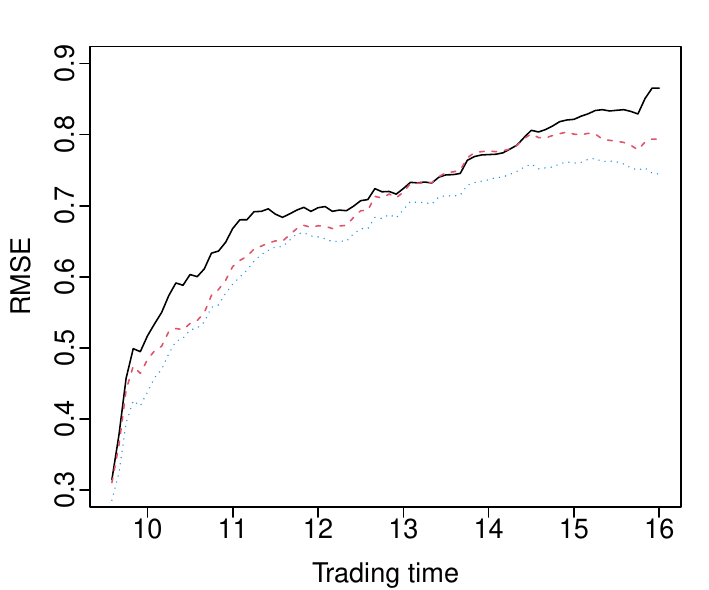}
\caption{Intraday $R^2$ and in-sample RMSE values between the observed and fitted CIDRs for BlackRock Inc.}\label{fig:3}
\end{figure}

In Figure~\ref{fig:3}, we compare the intraday $R^2$ and RMSE between the observed and fitted CIDRs for BlackRock Inc. Among the models, the PFLM demonstrates the highest $R^2$ values and the lowest RMSE values across intraday intervals, highlighting its strong in-sample fit. This performance is achieved through a penalization mechanism that imposes a smoothness constraint on the estimated regression coefficient surface, effectively controlling for potential overfitting.

The $R^2$ values for the PFLM model reach their highest levels between 10:30 am and 12:00 pm, indicating that the model captures a significant portion of intraday variability during this period of heightened trading activity. Following this peak, $R^2$ remains relatively stable until around 2:00 pm, after which it gradually declines in the final trading hours. This trend suggests that the PFLM’s penalized structure effectively captures and maintains the predictor-response relationship during periods of high liquidity and trading volume, while also preventing overfitting during lower-activity periods typically observed in the afternoon.

The $R^2$ values for the PFLM model reach their highest levels between 10:30 am and 12:00 pm, indicating that the model captures a significant portion of intraday variability during this period of heightened trading activity. Following this peak, $R^2$ remains relatively stable until around 2:00 pm, gradually declining in the final trading hours. This trend suggests that the PFLM’s penalized structure effectively captures and maintains the predictor-response relationship during high liquidity and trading volume while also preventing overfitting during lower-activity periods typically observed in the afternoon.

The superior performance of the PFLM model is due to its adaptive control over model complexity. Its smoothing penalty minimizes reactions to random noise in high-frequency data, resulting in stable and consistently high $R^2$ values throughout the day. In contrast, FPCR and FPLSR lack this penalization feature, making them more sensitive to fluctuations, especially during quieter trading hours, when market noise can disproportionately impact model predictions.

Averaged $R^2$ and RMSE are useful if one requires single numerical measures of fit. They can be expressed as
\begin{align*}
R^{2} &=\int_{\mathcal{I}} R^{2}(v) dv = \frac{1}{78}\sum^{78}_{i=1}R^{2}(v_i) \\
\text{RMSE} &= \int_{\mathcal{I}} \text{RMSE}(v)dv.
\end{align*}
The larger the $R^2$ value, the better the functional CAPM can capture the overall linear relationship between predictor and response. The smaller RMSE often reflects the larger $R^2$ value. 

We employ a linearity test between the decentered functional response and decentered functional predictor as proposed by \cite{EJGW2021} to examine the linear relationship between the intraday returns of an asset, denoted as $\bm{\Y}^{j,c}(v)$, and the stock $\bm{\X}^{c}(u)$. The null hypothesis is formulated as follows:
\begin{equation*}
H_0: m_{\beta} \in \mathcal{L} =  \left\lbrace m_{\beta}(\bm{\X}^{c})(v) = \int_{\mathcal{I}} \beta^j(u,v) \bm{\X}^{c}(u) du: \quad \beta^j \in \mathcal{L}^2[u \times v] \right\rbrace,
\end{equation*}
where $m_{\beta}$ is a Hilbert-Schmidt operator between $\mathcal{L}^2$ spaces that can be represented integrally using a bivariate kernel $\beta$, that is $\bm{\Y}^{j,c} = m_{\beta}(\bm{\X}^{c}) + \epsilon^j$. This test defines $H_0$ through an integral regression operator obtained by projecting the functional covariate and response into finite-dimensional functional directions. The Cram\'{e}r–von Mises statistic, which integrates these directions, measures the deviation of the empirical process from its expected zero mean. The statistic is calibrated using an efficient wild bootstrap on the residuals.

The results of the linearity test, conducted at a 0.05 significance level, reveal an apparent linear relationship for 449 firms between $\bm{\Y}^{j,c}(v)$ and $\bm{\X}^{c}(u)$. This indicates that our functional CAPM can adequately explain the intraday returns of these firms' assets. Conversely, for the remaining 58 firms, we observe $p$-values close to zero (i.e., $p \text{-value} < 0.05$), suggesting a potential nonlinear relationship between these firms' intraday returns and the stock. This implies that a more complex model may be required to capture the relationship for this subset of firms accurately.

\subsection{Out-of-sample prediction accuracy}\label{sec:5.2}

Apart from in-sample estimation accuracy, we compare one-step-ahead out-of-sample prediction accuracy among the methods by computing the RMSPE. The total RMSPE is defined as follows:
\begin{equation*}
\text{RMSPE} = \int_{\mathcal{I}} \text{RMSPE}(v)dv,
\end{equation*}
where
\begin{equation*}
\text{RMSPE}(v) = \sqrt{\frac{1}{n_{\text{test}}}\sum^{n_{\text{test}}}_{\zeta=1}[\Y_{\zeta}^j(v) - \widehat{\Y}_{\zeta}^{j}(v)]^2},  
\end{equation*}
where $\widehat{\Y}_{\zeta}^{j}(v)$ is the predicted response function. 

We consider an expanding-window approach to compare the out-of-sample prediction performance of the methods. For both datasets, we divide $n=251$ into two parts: a training sample consisting of first $n_{\text{train}} = 200$ curves and a test sample consisting of the remaining $n_{\text{test}} = 51$ curves. We obtain the one-step-ahead forecast for the $201\textsuperscript{st}$ curve using the entire observations in the training sets. We increase the training sample by one day and obtain the forecast for the $202\textsuperscript{st}$ curve. This procedure is repeated until the training samples cover the entire dataset.

The input for the traditional CAPM estimation is the daily closing price, and beta is a real value. In contrast, the parameters in the FPCR, FPLSR, and PFLM methods are estimated using the intraday data, and beta is a two-dimensional surface. Intuitively, the outcomes of the four methods are not directly comparable due to the nature of the data. Hence, we have performed an integration for the functional methods so that the mean and median of the four performance measures for the traditional CAPM, FPCR, FPLSR, and PFLM are comparable. 
\begin{table}[!htb]
\centering
\caption{Computed total $R^2$ and total RMSE values between the traditional classical CAPM and functional CAPM sorted by the 11 sectors covering $N = 488$ individual firms.}\label{tab:3}
\tabcolsep 0.054in
\begin{tabular}{@{}lrrrrrrrr@{}}
\toprule
 		&   \multicolumn{4}{c}{Total $R^2$} & \multicolumn{4}{c}{Total RMSE} \\\cmidrule{2-9}
Sector 				& CAPM & FPCR & FPLSR & PFLM & CAPM & FPCR & FPLSR & PFLM \\
\midrule
Energy (21) & 0.160 & 0.093 & 0.118 & 0.172 & 2.400 & 1.591 & 1.570 & 1.525 \\
Materials (27) & 0.207 & 0.139 & 0.180 & 0.222 & 1.813 & 1.117 & 1.092 & 1.066 \\
Industrials (69) & 0.234 & 0.154 & 0.205 & 0.246 & 1.897 & 0.993 & 0.963 & 0.939 \\
Consumer Discretionary (59) & 0.202 & 0.136 & 0.180 & 0.220 & 2.028 & 1.292 & 1.258 & 1.228 \\
Consumer Staples (31) & 0.105 & 0.060 & 0.111 & 0.145 & 1.223 & 0.843 & 0.820 & 0.804 \\
Health Care (61) & 0.149 & 0.111 & 0.148 & 0.189 & 1.678 & 1.042 & 1.020 & 0.997 \\ 
Financials (66) & 0.247 & 0.186 & 0.231 & 0.268 & 2.013 & 1.001 & 0.973 & 0.952 \\ 
Information Technology (76) & 0.254 & 0.192 & 0.253 & 0.300 & 2.454 & 1.090 & 1.047 & 1.015 \\
Communication Services (20) & 0.141 & 0.115 & 0.157 & 0.202 & 2.159 & 1.140 & 1.109 & 1.082 \\
Utilities (28) & 0.079 & 0.044 & 0.087 & 0.131 & 1.211 & 0.852 & 0.832 & 0.812 \\
Real Estate (30) & 0.160 & 0.070 & 0.121 & 0.151 & 1.630 & 0.987 & 0.960 & 0.944 \\
\midrule
Mean 	& 0.176 & 0.118 & 0.163 & 0.204 & 1.864 & 1.086 & 1.059 & 1.033 \\
Median & 0.160 & 0.115 & 0.157 & 0.202 & 1.897 & 1.042 & 1.020 & 0.997 \\
\bottomrule
\end{tabular}
\end{table}

Table~\ref{tab:3} displays the in-sample estimations using the total $R^2$ and RMSE values between the holdout and estimated responses. For the stocks in the S\&P 500 index, we summarize the results by their GICS sectors. The results presented in Table~\ref{tab:3} indicate that the model goodness-of-fit (measured by $R^2$) varies greatly amongst different industries. However, the rankings of the model fitness produced by the four methods are consistent. The traditional CAPM produces higher $R^2$ values than the FPCR and the FPLSR methods but underperforms the PFLM method. Information Technology, Financials, and Industrials are the sectors with the highest $R^2$ values, while Utilities and Consumer Staples have the lowest goodness of fit. This is consistent with existing literature that finds that stocks in some industries have less noisy prices and reflect relatively more firm-specific private information. Hence, this is reflected in the better goodness of fit for firms in these particular industries \citep[see, e.g.,][]{BNPW22}.

When investigating in-sample fitting (measured by RMSE), the PFLM method again outperforms the comparable methods with the smallest mean and median estimation errors. The sectors with better in-sample goodness of fit are Consumer Staples and Utilities, while the Energy and Consumer Discretionary sectors have relatively larger in-sample prediction errors. Overall, the PFLM method presents superiority with the accuracy of in-sample estimation. 

\begin{figure}[!htb]
\centering
\includegraphics[width=8.46cm]{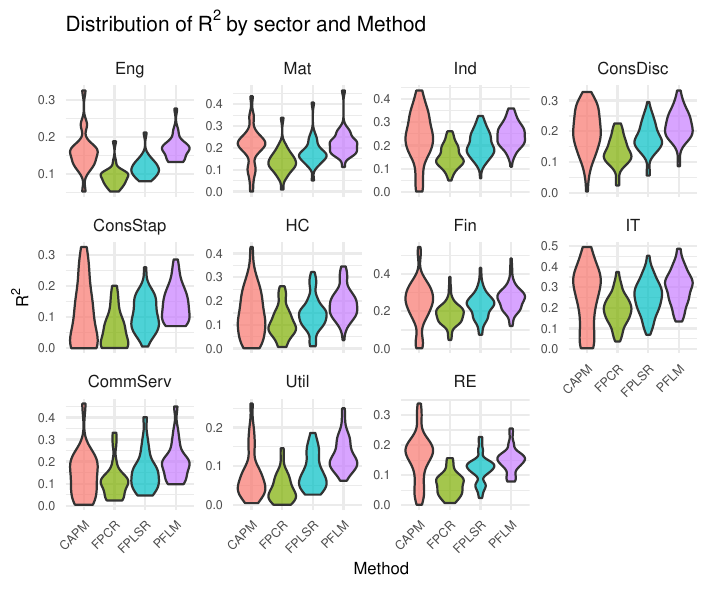}
\quad
\includegraphics[width=8.46cm]{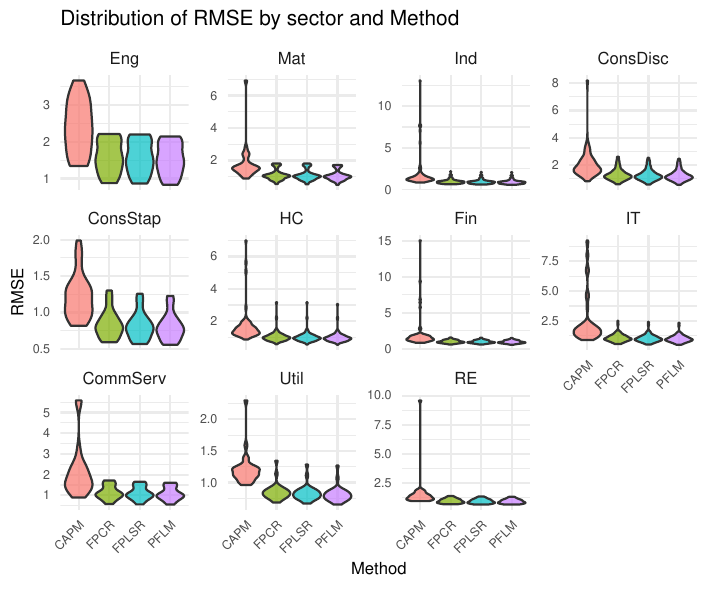}
\caption{Violin plots of $R^2$ and RMSE values obtained from the classical CAPM and functional CAPM with different estimation methods according to various sectors.}\label{fig:R_square_RMSE}
\end{figure}

As a graphical display, we present violin plots in Figure~\ref{fig:R_square_RMSE} to compare $R^2$ and RMSE values obtained from the CAPM and functional CAPM with different estimation methods for various industry sectors.

Table~\ref{tab:32} displays the model comparison on out-of-sample forecast accuracy using the total RMSPE values. It suggests that FPCR is the best performer with the smallest RMSPE, and PFLM is ranked second. The traditional CAPM is the worst performer, with the highest forecast errors. This observation is consistent regardless of the stock's GICS sector. The results further confirm the wider applicability of the FPCR for out-of-sample forecasting \citep[see, e.g.,][for a comparison between the FPCR and FPLSR]{WC23}.
\begin{table}[!htb]
\centering
\tabcolsep 0.35in
\caption{Computed total RMSPE values between the traditional classical CAPM and functional CAPM sorted by the 11 sectors covering $N = 488$ individual firms.}\label{tab:32}
\begin{tabular}{@{}lrrrr@{}}
\toprule
 		                    & \multicolumn{4}{c}{Total RMSPE}  \\
Sector 				        & CAPM & FPCR & FPLSR & PFLM \\
\midrule
Energy (21)                 & 1.966 & 1.228 & 1.466 & 1.265 \\ 
Materials (27)              & 1.536 & 0.917 & 1.130 & 0.938 \\ 
Industrials (69)            & 2.102 & 0.867 & 1.230 & 0.878 \\ 
Consumer Discretionary (59) & 2.332 & 1.134 & 1.572 & 1.139 \\ 
Consumer Staples (31)       & 1.287 & 0.732 & 1.186 & 0.753 \\ 
Health Care (61)            & 1.855 & 0.918 & 1.225 & 0.931 \\
Financials (66)             & 2.768 & 0.832 & 1.097 & 0.837 \\
Information Technology (76) & 3.485 & 0.982 & 1.466 & 0.985 \\ 
Communication Services (20) & 2.911 & 1.026 & 1.459 & 1.034 \\ 
Utilities (28)              & 1.135 & 0.669 & 1.063 & 0.669 \\ 
Real Estate (30)            & 2.064 & 0.855 & 1.333 & 0.875 \\ 
\midrule
Mean   & 2.131 & 0.924 & 1.293 & 0.937 \\
Median & 2.064 & 0.917 & 1.230 & 0.931 \\
\bottomrule
\end{tabular}
\end{table}

\begin{figure}[!htb]
\centering
\includegraphics[width=13cm]{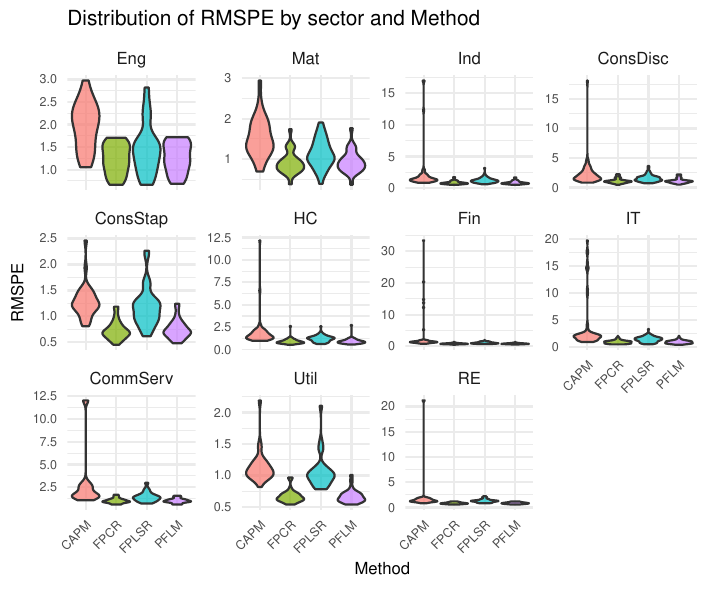}
\caption{Violin plots of RMSPE values were obtained from the classical CAPM and functional CAPM using different estimation methods for various sectors.}\label{fig:RMSPE}
\end{figure}

As a graphical display, we present violin plots in Figure~\ref{fig:RMSPE} to compare RMSPE values obtained from the CAPM and functional CAPM with different estimation methods according to various industry sectors.

The PFLM method performs best in model goodness-of-fit and in-sample estimation accuracy, while FPCR performs best in out-of-sample forecast. The above-mentioned functional methods present superior performance than the traditional CAPM. This is because the traditional CAPM uses only one observation for a trading day (closing price recorded at 16:00); it is highly aggregated and ignores the rich information embedded in the intraday fluctuation. On the other hand, the functional methods utilize intraday data  (78 data points) for a single trading day. The functional methods utilize $B$-spline basis functions to reconstruct smooth curves from discrete data. An adequate amount of smoothing mitigates the data measurement error often ignored by the traditional CAPM.

To assess the robustness of our findings from the 2021 data, we conducted a sensitivity analysis in Appendix~\ref{sec:Appendix_A} by replicating the study using high-frequency intraday data from 2020.

\subsection{Firm characteristics and model performance}\label{sec:5.3}

We compare the model performance measures among the four methods and investigate whether firm characteristics impact these measures. We sort each of the S\&P 500 firms by firm characteristics, then adopt the two-sample $t$ tests to investigate whether model performance (measured by $R^2$, RMSE, and RMSPE) is significantly different between the top and bottom decile sub-sample groups. We follow \cite{BNPW22} and investigate the following firm characteristics: log market capitalization ($\ln MC_{i,t}$), log stock price ($\ln P_{i,t}$), and leverage ($LEV_{i,t}$). We follow \cite{K09} and adopt volume turnover ($VOL_{i,t}$) to proxy valuation uncertainty, which is measured as the ratio of the number of shares traded in a month and the number of shares outstanding. Next, we include two measures of illiquidity and trading costs to capture limits to arbitrage. The first is the illiquidity measure ($ILLIQ_{i,t}$) in \cite{A02} and \cite{BCT09}, which is defined as the average ratio of the daily absolute return to the (dollar) trading volume on that day. The second is the stock's average effective bid-ask spread ($BidAskSpread_{i,t}$), which is measured by the natural logarithm of the average daily effective spread.

Table~\ref{tab:firm_character} presents the results of mean differences for model performance when comparing firms with different characteristics. For model goodness-of-fit, as measured by $R^2$, all four methods suggest significantly superior performance for larger ($\ln MC_{i,t}$) and higher-priced ($\ln P_{i,t}$) stocks. This is consistent with prior studies in the literature showing that larger firms are more transparent and liquid and are, therefore, more efficiently priced \citep{BNPW22}. The three functional methods yield significantly better model goodness-of-fit for stocks with higher illiquidity ($ILLIQ_{i,t}$), and FPCR and PFLM perform better for firms with higher bid-ask spread ($BidAskSpread_{i,t}$). It indicates that the functional methods provide superior model goodness-of-fit for less liquid stocks with greater trading costs than the traditional CAPM.
\begin{center}
\begin{footnotesize}
\tabcolsep 0.025in
\renewcommand*{\arraystretch}{0.85}
\begin{longtable}{@{}llcccccccccccc@{}}
\caption{The impact of firm characteristics on model performance.}\label{tab:firm_character} \\
\toprule
Firm & & \multicolumn{4}{c}{$R^2$} & \multicolumn{4}{c}{RMSE} & \multicolumn{4}{c}{RMSPE} \\ 
\multicolumn{2}{l}{\hspace{-.04in}{characteristic}} & CAPM & FPCR & FPLSR & PFLM & CAPM & FPCR & FPLSR & PFLM & CAPM & FPCR & FPLSR & PFLM \\ \midrule
\endfirsthead
\toprule
Firm & & \multicolumn{4}{c}{$R^2$} & \multicolumn{4}{c}{RMSE} & \multicolumn{4}{c}{RMSPE} \\ 
\multicolumn{2}{l}{\hspace{-.04in}{characteristics}} & CAPM & FPCR & FPLSR & PFLM & CAPM & FPCR & FPLSR & PFLM & CAPM & FPCR & FPLSR & PFLM \\ \midrule
\endhead
\hline \multicolumn{14}{r}{{Continued on next page}} \\ 
\endfoot
\hline  
\endlastfoot
$\ln MC_{i,t}$ & High & 0.227 & 0.167 & 0.212 & 0.260 & 1.494 & 0.869 & 0.845 & 0.819 & 1.554 & 0.800 & 1.051 & 0.798 \\ 
& Low & 0.152 & 0.108 & 0.152 & 0.191 & 2.324 & 1.324 & 1.289 & 1.261 & 2.881 & 1.118 & 1.644 & 1.137 \\ 
& t.statistic & 3.208 & 3.676 & 3.323 & 3.901 & -3.059 & -6.246 & -6.248 & -6.383 & -2.699 & -4.555 & -5.239 & -4.911 \\ 
& p.value & 0.002 & 0.001 & 0.001 & 0.000 & 0.003 & 0.000 & 0.000 & 0.000 & 0.010 & 0.000 & 0.000 & 0.000 \\ 
& Sig. & *** & *** & *** & *** & ** & *** & *** & *** & *** & *** & *** & *** \\
\cmidrule{2-14}
$\ln P_{i,t}$ & High & 0.235 & 0.173 & 0.220 & 0.268 & 2.173 & 1.064 & 1.032 & 1.002 & 3.347 & 0.962 & 1.371 & 0.960 \\ 
& Low & 0.160 & 0.108 & 0.148 & 0.188 & 2.094 & 1.344 & 1.314 & 1.284 & 2.145 & 1.087 & 1.530 & 1.110 \\ 
& t.statistic & 3.361 & 4.576 & 4.470 & 5.066 & 0.269 & -3.903 & -3.990 & -4.085 & 1.762 & -1.949 & -1.511 & -2.343 \\ 
& p.value & 0.001 & 0.000 & 0.000 & 0.000 & 0.789 & 0.000 & 0.000 & 0.000 & 0.083 & 0.054 & 0.134 & 0.021 \\  
& Sig. & *** & *** & *** & ***  & & *** & *** & *** &  & & & * \\ 
\cmidrule{2-14}
$LEV_{i,t}$ & High & 0.192 & 0.135 & 0.182 & 0.220 & 2.091 & 1.081 & 1.052 & 1.028 & 2.939 & 0.958 & 1.339 & 0.970 \\ 
& Low & 0.221 & 0.160 & 0.214 & 0.258 & 2.591 & 1.129 & 1.093 & 1.061 & 3.518 & 0.993 & 1.411 & 0.996 \\ 
& t.statistic & -0.331 & -1.093 & -0.722 & -0.943 & 0.665 & 0.616 & 0.588 & 0.601 & 0.347 & -0.146 & 1.429 & 0.037 \\ 
& p.value & 0.742 & 0.278 & 0.472 & 0.348 & 0.508 & 0.540 & 0.558 & 0.550 & 0.730 & 0.884 & 0.157 & 0.971 \\ 
& Sig. & \\ 
\cmidrule{2-14}  
$VOL_{i,t}$ & High & 0.164 & 0.110 & 0.159 & 0.201 & 1.726 & 1.048 & 1.017 & 0.992 & 1.998 & 0.890 & 1.302 & 0.901 \\ 
& Low & 0.187 & 0.136 & 0.182 & 0.226 & 2.352 & 1.035 & 1.006 & 0.979 & 3.288 & 0.907 & 1.294 & 0.920 \\ 
& t.statistic & -0.932 & -0.799 & -1.153 & -0.880 & -0.107 & 3.236 & 3.273 & 3.254 & -2.423 & 2.849 & 1.227 & 2.943 \\ 
& p.value & 0.354 & 0.426 & 0.252 & 0.381 & 0.915 & 0.002 & 0.002 & 0.002 & 0.019 & 0.006 & 0.224 & 0.005 \\ 
& Sig. & & & & & & *** & *** & *** & * & ** & & ** \\ 
\cmidrule{2-14}
$ILLIQ_{i,t}$ & High & 0.185 & 0.126 & 0.169 & 0.212 & 1.848 & 1.047 & 1.021 & 0.995 & 1.879 & 0.868 & 1.208 & 0.881 \\ 
& Low & 0.152 & 0.113 & 0.161 & 0.202 & 2.286 & 1.140 & 1.107 & 1.081 & 3.222 & 0.999 & 1.417 & 1.010 \\ 
& t.statistic & -1.654 & -2.558 & -2.355 & -3.005 & 2.151 & 4.807 & 4.786 & 5.011 & 1.940 & 2.402 & 3.238 & 2.795 \\ 
& p.value & 0.102 & 0.012 & 0.021 & 0.004 & 0.035 & 0.000 & 0.000 & 0.000 & 0.058 & 0.018 & 0.002 & 0.006 \\ 
& Sig. & & * & * & ** & ** & *** & *** & *** &  & * & ** & ** \\ 
\cmidrule{2-14}
$BidAskSpread_{i,t}$ & High & 0.216 & 0.135 & 0.187 & 0.226 & 1.539 & 0.983 & 0.952 & 0.930 & 1.796 & 0.850 & 1.259 & 0.864 \\ 
& Low & 0.193 & 0.133 & 0.177 & 0.220 & 1.719 & 1.046 & 1.018 & 0.992 & 2.119 & 0.895 & 1.262 & 0.901 \\ 
& t.statistic & 0.938 & 2.400 & 1.679 & 2.040 & 4.499 & 6.562 & 6.483 & 6.471 & 3.170 & 6.317 & 5.782 & 6.140 \\ 
& p.value & 0.351 & 0.018 & 0.096 & 0.044 & 0.000 & 0.000 & 0.000 & 0.000 & 0.003 & 0.000 & 0.000 & 0.000 \\ 
& Sig. & & * & & * & *** & *** & *** & *** & ** & *** & *** & ***\\
\bottomrule
\end{longtable}
 \end{footnotesize}
 \vspace{-.2in}
 \begin{justify}
\footnotesize{*** significance at 0.001, ** significance at 0.01, * significance at 0.05\\
Definitions of firm characteristics are defined: $\ln \text{MC}_{i,t}$ is the log market capitalization for firm $i$ at time $t$ and $\ln \text{P}_{i,t}$ is the log price. $\text{LEV}_{i,t}$ is leverage measured by the ratio of total liabilities to total assets. $\text{VOL}_{i,t}$ monthly volume turnover is a valuation uncertainty proxy, which is measured as the ratio of the number of shares traded in a month and the number of shares outstanding. $\text{ILLIQ}_{i,t}$ is the \cite{A02} measure of illiquidity, which is measured by the average ratio of the daily absolute return to the (dollar) trading volume on that day. $\text{BidAskSpread}_{i,t}$ is the average effective bid-ask spread, which is measured by the natural logarithm of the average daily effective spread.}
\end{justify}
\end{center}  

When measuring in-sample fitting, all four methods produce significantly smaller RMSE for stocks with greater market capitalization ($\ln MC_{i,t}$), higher illiquidity ($ILLIQ_{i,t}$), and greater trading costs ($BidAskSpread_{i,t}$). It is also evident that the three functional methods perform significantly better for higher-priced stocks ($\ln P_{i,t}$) and stocks with lower monthly volume turnover ($VOL_{i,t}$). At the same time, the performance difference for the traditional CAPM was not observed for such stocks. The test results for the out-of-sample forecast accuracy are similar. The RMSPE is significantly smaller for stocks with greater market capitalization ($\ln MC_{i,t}$) and greater trading costs ($BidAskSpread_{i,t}$). Similarly, the three functional methods perform significantly better for firms with higher illiquidity ($ILLIQ_{i,t}$), while the performance difference for the traditional CAPM was not observed.

\subsection{Corporate governance and model performance}\label{sec:5.4}

We also investigate model performance due to external monitoring and corporate governance factors. We adopt analyst-following variables and institutional holdings to measure the external monitoring effect. Financial analysts gather information from diverse sources, assess current firm performance, forecast future prospects, and provide buy, hold, or sell recommendations to investors. Prior literature indicates analysts mitigate information asymmetry across various dimensions, acting as external monitors for firm managers. Consequently, they impact firms' investment, financing decisions, stock prices, liquidity, and valuation \citep{HT13, HLS00, DK13}. To investigate the effect of analysts following on model performance, we adopt the widely used analyst coverage ($\text{Coverage}_{i,t}$) \citep{HT13} and analyst forecast error ($\text{FE}_{i,t}$) \citep{P08} measures. The $\text{Coverage}_{i,t}$ is measured as:
\begin{equation*}
\text{Coverage}_{i,t} = \ln(1+\text{No.Forecasts}_{i,t}),
\end{equation*}
where $\text{No.Forecasts}_{i,t}$ is the average of the 12 monthly numbers of earnings forecasts for firm $i$ in year $t$, and the numbers of forecasts are obtained from the $I/B/E/S$ summary file. A larger $\text{Coverage}_{i,t}$ indicates stronger external monitoring.

Following \cite{P08}, we measure analyst forecast error ($FE_{i,t}$) as:
\begin{equation*}
\text{FE}_{i,t} = \frac{|\text{Forecast}_{i,t} - \text{Actual}_{i,t}|}{P_{i,t}},
\end{equation*}
where $\text{Forecasts}_{i,t}$ is the mean of the forecasts made by each analyst for firm $i$ in year $t$ before the client's earnings announcement date, and $\text{Actual}_{i,t}$ is firm's actual earnings in year $t$. The absolute value of the forecast error is deflated by the latest available stock price. A smaller forecast error in the measurement ($\text{FE}_{i,t}$) represents a greater forecast accuracy.

Institutional holding ($\text{InstoHold}_{i,t}$) is measured as the percentage of outstanding shares held by institutional investors. It is commonly acknowledged that institutional investors are informed traders in markets responsible for impounding price information and contribute to market efficiency through their trading \citep{BK009}. Firms under greater external monitoring effect are expected to provide higher quality firm-specific information to enable better model goodness-of-fit and smaller forecast errors. We adopt two additional board structure variables to measure effective corporate governance. Following \cite{LNY08}, we use board independence measured by the proportion of outside independent directors ($\text{Independent}_{i,t}$), and board leadership by a dummy variable of one if the CEO is the chair of the board ($\text{Duality}_{i,t}$).

Table~\ref{tab:character} presents the results of mean differences for model performance when comparing firms with different levels of external monitoring and corporate governance. For model goodness-of-fit, all three methods perform significantly better for firms with higher analyst coverage ($\text{Coverage}_{i,t}$) and greater analyst forecast accuracy (i.e., lower $\text{FE}_{i,t}$). This is consistent with studies that find that firms with greater analyst following produce more transparent firm-specific information, which improves price efficiency. In addition, the three functional methods provide greater $R^2$ for firms with a lower proportion of independent directors, while no difference is observed for the traditional CAPM method. For in-sample estimation, all four methods provide significantly smaller estimation errors for stocks with higher analyst forecast accuracy, and the three functional methods perform better for firms with higher institutional holding ($\text{InstoHold}_{i,t}$). For out-of-sample prediction, when using the functional methods, RMSPE is significantly smaller for firms with better analyst forecast accuracy (lower $\text{FE}_{i,t}$) and higher institutional holding ($\text{InstoHold}_{i,t}$). Consistent with prior findings, the three functional methods produce much smaller RMSE than the traditional CAPM method for in-sample estimation and out-of-sample prediction.
\begin{center}
\begin{footnotesize}
\tabcolsep 0.03in
\renewcommand*{\arraystretch}{0.85}
\begin{longtable}{@{}llcccccccccccc@{}}
\caption{The impact of analyst following and corporate governance on model performance.}\label{tab:character} \\
\toprule
\multicolumn{2}{l}{\hspace{-.05in}{Analyst following \&}} & \multicolumn{4}{c}{$R^2$} & \multicolumn{4}{c}{RMSE} & \multicolumn{4}{c}{RMSPE} \\ 
\multicolumn{2}{l}{\hspace{-.05in}{corporate governance}} & CAPM & FPCR & FPLSR & PFLM & CAPM & FPCR & FPLSR & PFLM & CAPM & FPCR & FPLSR & PFLM\\ \midrule
\endfirsthead
\toprule
\multicolumn{2}{l}{\hspace{-.05in}{Analyst following \&}} & \multicolumn{4}{c}{$R^2$} & \multicolumn{4}{c}{RMSE} & \multicolumn{4}{c}{RMSPE} \\ 
\multicolumn{2}{l}{\hspace{-.05in}{corporate governance}} & CAPM & FPCR & FPLSR & PFLM & CAPM & FPCR & FPLSR & PFLM & CAPM & FPCR & FPLSR & PFLM\\ \midrule
\endhead
\midrule \multicolumn{14}{r}{{Continued on next page}} \\ 
\endfoot
\hline
\endlastfoot
$Coverage_{i,t}$ & High & 0.259 & 0.176 & 0.225 & 0.270 & 1.816 & 1.052 & 1.021 & 0.991 & 1.806 & 0.964 & 1.328 & 0.954 \\ 
& Low & 0.170 & 0.106 & 0.151 & 0.188 & 1.938 & 1.064 & 1.037 & 1.015 & 2.638 & 0.928 & 1.362 & 0.943 \\ 
& t.statistic & 4.185 & 5.366 & 5.056 & 5.393 & -0.351 & -0.200 & -0.275 & -0.413 & -1.227 & 0.622 & -0.388 & 0.197 \\ 
& p.value & 0.000 & 0.000 & 0.000 & 0.000 & 0.727 & 0.842 & 0.784 & 0.680 & 0.225 & 0.536 & 0.699 & 0.844 \\ 
& Sig. & *** & *** & *** & *** & & & & &\\
\cmidrule{2-14}
$FE_{i,t}$ & High & 0.185 & 0.125 & 0.171 & 0.211 & 2.195 & 1.160 & 1.131 & 1.104 & 2.059 & 0.976 & 1.409 & 0.998 \\ 
& Low & 0.210 & 0.142 & 0.190 & 0.233 & 2.100 & 1.105 & 1.074 & 1.047 & 2.859 & 0.966 & 1.333 & 0.975 \\ 
& t.statistic & -2.690 & -3.665 & -3.154 & -3.604 & 2.298 & 5.695 & 5.649 & 5.716 & 2.068 & 4.484 & 4.284 & 4.712 \\ 
& p.value & 0.009 & 0.000 & 0.002 & 0.001 & 0.024 & 0.000 & 0.000 & 0.000 & 0.041 & 0.000 & 0.000 & 0.000 \\ 
& Sig. & ** & *** & ** & *** & * & *** & *** & *** & * & *** & *** & *** \\
\cmidrule{2-14}
$InstoHold_{i,t}$ & High & 0.188 & 0.132 & 0.187 & 0.223 & 1.995 & 1.065 & 1.030 & 1.009 & 2.694 & 0.908 & 1.339 & 0.920 \\ 
& Low & 0.167 & 0.117 & 0.160 & 0.201 & 1.929 & 1.209 & 1.182 & 1.153 & 2.162 & 0.998 & 1.322 & 1.013 \\ 
& t.statistic & 0.502 & 1.885 & 0.681 & 0.844 & 4.316 & 6.615 & 6.471 & 6.453 & 2.260 & 5.939 & 2.450 & 5.573 \\ 
& p.value & 0.660 & 0.162 & 0.559 & 0.475 & 0.000 & 0.000 & 0.000 & 0.000 & 0.029 & 0.000 & 0.075 & 0.000 \\ 
& Sig. & & & & & *** & *** & *** & *** & * & *** & & *** \\
\cmidrule{2-14}
$Independent_{i,t}$ & High & 0.196 & 0.116 & 0.158 & 0.202 & 1.816 & 1.074 & 1.049 & 1.023 & 1.816 & 0.890 & 1.245 & 0.910 \\ 
& Low & 0.210 & 0.147 & 0.192 & 0.229 & 1.839 & 1.186 & 1.157 & 1.131 & 2.055 & 0.990 & 1.391 & 1.004 \\ 
& t.statistic & -1.067 & -2.556 & -2.546 & -2.806 & -1.070 & -0.851 & -0.755 & -0.715 & -1.039 & -1.702 & -1.362 & -1.232 \\ 
& p.value & 0.289 & 0.013 & 0.013 & 0.006 & 0.288 & 0.398 & 0.453 & 0.477 & 0.302 & 0.093 & 0.177 & 0.222 \\ 
& Sig. & & * & * & ** & & & & & \\
$Duality_{i,t}$ & High & 0.196 & 0.134 & 0.181 & 0.222 & 1.946 & 1.080 & 1.050 & 1.024 & 2.306 & 0.924 & 1.315 & 0.935 \\ 
& Low & 0.188 & 0.127 & 0.173 & 0.212 & 1.825 & 1.102 & 1.073 & 1.049 & 2.178 & 0.941 & 1.317 & 0.958 \\ 
& t.statistic & 0.558 & 0.769 & 0.837 & 1.082 & 0.819 & -0.515 & -0.543 & -0.602 & 0.397 & -0.478 & -0.027 & -0.613 \\ 
& p.value & 0.578 & 0.443 & 0.404 & 0.281 & 0.414 & 0.607 & 0.588 & 0.549 & 0.692 & 0.634 & 0.978 & 0.541 \\
& Sig. & & & & & & & & & \\
\bottomrule
\end{longtable}
\end{footnotesize}
\vspace{-.2in}
\begin{justify}
\footnotesize{*** significance at 0.001, ** significance at 0.01, * significance at 0.05\\
Definitions of firm characteristics: $\text{Coverage}_{i,t}$ is the natural logarithm of one plus the average of 12 monthly numbers of earnings forecasts. $\text{FE}_{i,t}$ is measured by the absolute analyst earnings forecast error deflated by the latest available stock price, where the forecast error is equal to the analyst consensus forecast minus actual earnings. $\text{InstoHold}_{i,t}$ is the percentage of outstanding shares institutional investors hold. $\text{Independent}_{i,t}$ is the proportion of outside independent directors. $\text{Duality}_{i,t}$ is a dummy variable that equals one if the CEO is the board chair.}
\end{justify}
\end{center}  

Furthermore, we examine the computational efficiency of each method by recording the elapsed time required to fit individual stock datasets. Since each method is applied independently to each dataset, we report the average computation time across all stocks. Our findings reveal that, on average, the CAPM, FPCR, FPLSR, and PFLM require 0.010, 3.729, 0.097, and 7.924 seconds, respectively, to fit a single dataset. As anticipated, the CAPM is the most computationally efficient method. FPLSR demonstrates greater computational efficiency among the functional regression models, requiring less time than both FPCR and PFLM. Notably, PFLM exhibits the highest computational cost, largely due to its use of a penalization algorithm and a grid search procedure for optimizing the regularization parameter, significantly increasing the overall computation time. We note that the computations were performed using R 4.2.2 on an Intel Core i7 6700HQ 2.6 GHz PC.

In summary, this study finds that the PFLM proposed here performs better in model goodness-of-fit and in-sample fitting than the traditional CAPM. At the same time, the FPCR method produces the best out-of-sample prediction among the comparable models. The results also suggest that the firm characteristics significantly influence the performance of functional models. The functional methods outperform the traditional CAPM for stocks in model goodness-of-fit with greater trading costs and illiquidity. Similar findings are evident in predictions that functional methods perform better for firms with lower monthly volume turnover, greater illiquidity, and greater bid-ask spread. Such firms are usually considered as more information opaque \citep{A02, BNPW22}, so that is where we see the use of higher-frequency data and functional methods as much superior for asset pricing, reinforcing their ability to capture firm-specific information and price dynamics more effectively than the traditional CAPM.

\section{Conclusion}\label{sec:6}

Given an ever-increasing amount of high-frequency financial data, we propose an extension of the CAPM in which the predictor and response are function-valued variables. With high-frequency data, the functional CAPM is a new approach to empirically test the classical model by studying the linear relationship between stocks and a market index. The functional approach can enable the risk that an individual stock contributes to the overall portfolio to be better managed. Overall, the functional CAPM estimation enables improved portfolio optimization \citep[see, e.g.,][]{HPS20, HPS20b, DHT+25}. We consider FPCR, FPLSR, and PFLM to estimate the bivariate regression coefficient function in this concurrent function-on-function linear regression. The estimated regression coefficient function measures a linear relationship between the functional predictor and response. We can obtain fitted responses with the estimated regression coefficient function and compare their values with the holdout ones. Via the intraday and total $R^2$ and RMSE, we evaluate and compare the goodness-of-fit between a market index and its constituents using the functional CAPM. Via the RMSPE, we also study its out-of-sample forecast accuracy, which can be divided into various GICS industry sectors. The findings in this study suggest that the PFLM, along with high-frequency data, presents superior in-sample goodness-of-fit, while the FPCR method performs best in terms of out-of-sample predictions, followed closely by the PFLM. The functional CAPM provides better model goodness-of-fit and prediction accuracy for less price-efficient or more information-opaque stocks than the traditional CAPM.

There are at least four ways in which the methodology presented can be extended. 
\begin{inparaenum}[1)]
\item We used one-year intraday data, but the analysis could have been conducted for a longer period. 
\item The functional CAPM is an example of the concurrent function-on-function linear model. Following \cite{Corsi09} and \cite{HPS20}, one extension is to add lagged variables of the response variable $\Y_{t-1}^j(v)$, $\Y_{t-5}^j(v)$ and $\Y_{t-22}^j(v)$ representing the past daily, weekly, and monthly CIDRs of the $j\textsuperscript{th}$ stock. 
\item Following \cite{QL19}, the other extension is to consider a \textit{nonlinear} function-on-function regression, where the beta surface can be estimated non-parametrically. These represent opportunities for further investigation. The functional CAPM assumes a linear relationship between the functional predictor and response. We validate this assumption by performing a linearity test, which shows that this linear structure is appropriate for over 400 stocks in our dataset. However, given the potential for nonlinear dependencies in financial data, particularly at high frequencies, it is important to acknowledge the limitations of this assumption. Our functional framework can be adapted if nonlinear relationships are present by utilizing alternative basis functions, such as polynomial or wavelet bases, to approximate nonlinear patterns. This flexibility allows us to extend the model to accommodate non-linearities as needed. 
\item Model misspecification due to unmodeled non-linearities could impact the functional CAPM's performance, potentially leading to biased estimates. Future work could explore nonlinear function-on-function regression models to account for these dynamics, enhancing the model's robustness in diverse market conditions.
\end{inparaenum}

\section*{Acknowledgment}

The authors are grateful for the insightful comments provided by the reviewers, which led to a much-improved manuscript. The authors also thank participants at the 2022 International Conference on Trends and Perspectives in Linear Statistical Inference in Tomar, Portugal, for their insightful comments and suggestions. The authors thank Associate Professor Yajun Xiao for the detailed discussion of this paper at the 2024 Sydney Banking and Financial Stability Conference. This work was supported by an Australian Research Council Discovery Project (grant no: DP230102250) and the Scientific and Technological Research Council of Turkey (TUBITAK) (grant no: 120F270).

\newpage
\appendix

\section{Sensitivity analysis}\label{sec:Appendix_A}

This additional analysis examines the consistency of the performance metrics across different periods and market conditions. Specifically, we recalculated the in-sample goodness-of-fit (measured by $R^2$ and RMSE)  and the out-of-sample forecast accuracy (measured by RMSPE) for the traditional daily CAPM and the three functional methods: FPCR, FPLSR, and PFLM. The results are summarized in Tables~\ref{tab:3_2020} and~\ref{tab:32_2020}. This sensitivity analysis allows us to evaluate whether the conclusions drawn for 2021 remain valid when applied to data from a different year, particularly a year with potentially different market dynamics and trading patterns.
\begin{table}[!htb]
\centering
\caption{Computed total $R^2$ and total RMSE values between the traditional daily CAPM and functional CAPM sorted by the 11 sectors for 2020 data covering $N = 488$ individual firms.}\label{tab:3_2020}
\tabcolsep 0.05in
\begin{tabular}{@{}lrrrrrrrr@{}}
\toprule
 		&   \multicolumn{4}{c}{Total $R^2$} & \multicolumn{4}{c}{Total RMSE} \\
Sector & CAPM & FPCR & FPLSR & PFLM & CAPM & FPCR & FPLSR & PFLM \\
\midrule
Energy (21) & 0.409 & 0.065 & 0.288 & 0.374 & 4.031 & 2.155 & 1.862 & 1.746 \\
Materials (27) & 0.543 & 0.074 & 0.303 & 0.399 & 2.318 & 1.448 & 1.237 & 1.155 \\
Industrials (69) & 0.438 & 0.065 & 0.280 & 0.378 & 7.616 & 2.262 & 2.050 & 1.956 \\
Consumer Discretionary (59) & 0.378 & 0.072 & 0.289 & 0.389 & 8.643 & 2.407 & 2.154 & 2.042 \\
Consumer Staples (31) & 0.380 & 0.068 & 0.297 & 0.393 & 2.943 & 1.282 & 1.116 & 1.049 \\
Health Care (61) & 0.391 & 0.060 & 0.274 & 0.365 & 6.345 & 2.005 & 1.812 & 1.733 \\ 
Financials (66) & 0.490 & 0.111 & 0.320 & 0.411 & 8.199 & 2.301 & 2.084 & 1.991 \\ 
Information Technology (76) & 0.378 & 0.078 & 0.291 & 0.389 & 10.881 & 2.795 & 2.561 & 2.465 \\
Communication Services (20) & 0.302 & 0.074 & 0.265 & 0.350 & 8.529 & 2.403 & 2.208 & 2.125 \\
Utilities (28) & 0.462 & 0.092 & 0.338 & 0.436 & 4.679 & 1.674 & 1.459 & 1.369 \\
Real Estate (30) & 0.449 & 0.068 & 0.278 & 0.363 & 9.325 & 2.458 & 2.240 & 2.154 \\
\midrule
Mean 	& 0.419 & 0.075 & 0.293 & 0.386 & 6.682 & 2.108 & 1.889 & 1.798 \\
Median & 0.409 & 0.072 & 0.288 & 0.388 & 7.616 & 2.262 & 2.049 & 1.956 \\
\bottomrule
\end{tabular}
\end{table}

Table~\ref{tab:3_2020} displays the in-sample estimations using the total $R^2$ and RMSE values between the holdout and estimated responses for the 2020 data. For the stocks in the S\&P 500 index, the results are summarized by their GICS sectors. The results indicate that the model goodness-of-fit (measured by $R^2$) varies considerably across different industries. However, the rankings of the model performance remain consistent across methods. The traditional CAPM generally produces higher R2R2 values than the FPCR and FPLSR methods but underperforms the PFLM method. Information Technology, Financials, and Industrials are the sectors with the highest $R^2$ values, while Utilities and Communication Services have the lowest goodness of fit. These findings are consistent with the literature, suggesting that stocks in sectors with greater price efficiency and firm-specific information tend to yield better model goodness-of-fit.

When examining the fit of the sample using the RMSE values, the PFLM method again demonstrates its superiority by achieving the smallest mean and median estimation errors. This observation is consistent across sectors, with PFLM showing particular strength in Consumer Staples and Utilities, sectors with relatively stable and predictable pricing. In contrast, sectors such as Information Technology and Consumer Discretionary exhibit larger RMSE values, which can be attributed to their higher price volatility and sensitivity to market dynamics. Overall, the PFLM method maintains its position as the best performer in terms of in-sample estimation accuracy.
\begin{table}[!htb]
\centering
\tabcolsep 0.35in
\caption{Computed total RMSPE values between the traditional classical CAPM and functional CAPM sorted by the 11 sectors for 2020 data covering $N = 488$ individual firms.}\label{tab:32_2020}
\begin{tabular}{@{}lrrrr@{}}
\toprule
 		                    & \multicolumn{4}{c}{Total RMSPE}  \\
Sector 				        & CAPM & FPCR & FPLSR & PFLM \\
\midrule
Energy (21)                 & 3.497 & 1.771 & 3.319 & 1.703 \\ 
Materials (27)              & 1.889 & 1.034 & 2.051 & 1.012 \\ 
Industrials (69)            & 4.768 & 2.189 & 2.803 & 2.159 \\ 
Consumer Discretionary (59) & 3.760 & 1.817 & 3.235 & 1.870 \\ 
Consumer Staples (31)       & 2.332 & 0.897 & 1.827 & 0.895 \\ 
Health Care (61)            & 4.178 & 1.642 & 2.408 & 1.655 \\
Financials (66)             & 5.652 & 2.182 & 2.811 & 2.202 \\
Information Technology (76) & 7.121 & 2.675 & 3.342 & 2.669 \\ 
Communication Services (20) & 5.176 & 1.997 & 2.688 & 2.034 \\ 
Utilities (28)              & 3.366 & 1.303 & 2.164 & 1.340 \\ 
Real Estate (30)            & 4.336 & 2.238 & 3.123 & 2.239 \\ 
\midrule
Mean   & 4.188 & 1.795 & 2.706 & 1.797 \\
Median & 4.177 & 1.816 & 2.803 & 1.870 \\
\bottomrule
\end{tabular}
\end{table}

Table~\ref{tab:32_2020} presents the results of out-of-sample forecast accuracy using RMSPE values. The findings indicate that the FPCR method provides the smallest RMSPE values across most sectors, confirming its robustness in forecasting performance. The PFLM method follows closely, with slightly higher RMSPE values, while the traditional CAPM remains the weakest performer with the largest forecast errors. This pattern is consistent regardless of the sector, further reinforcing the advantages of functional methods for out-of-sample predictions. Notably, Utilities and Consumer Staples show the lowest RMSPE values across all methods, indicating better predictability in these sectors. In contrast, Information Technology and Financials exhibit the highest RMSPE values, likely reflecting greater volatility and complexity in their price dynamics.

\begin{figure}[!htb]
\centering
\subfloat[$R^2$ value]
{\includegraphics[width=8.46cm]{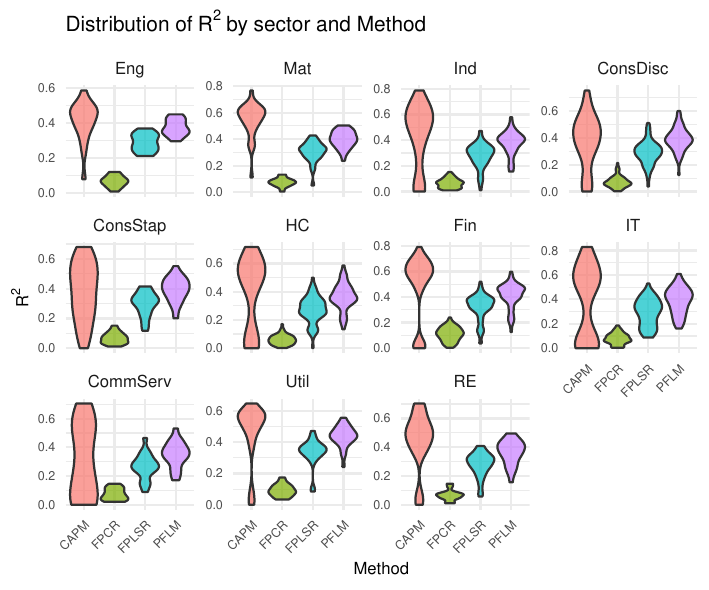}}
\quad
\subfloat[RMSE value]
{\includegraphics[width=8.46cm]{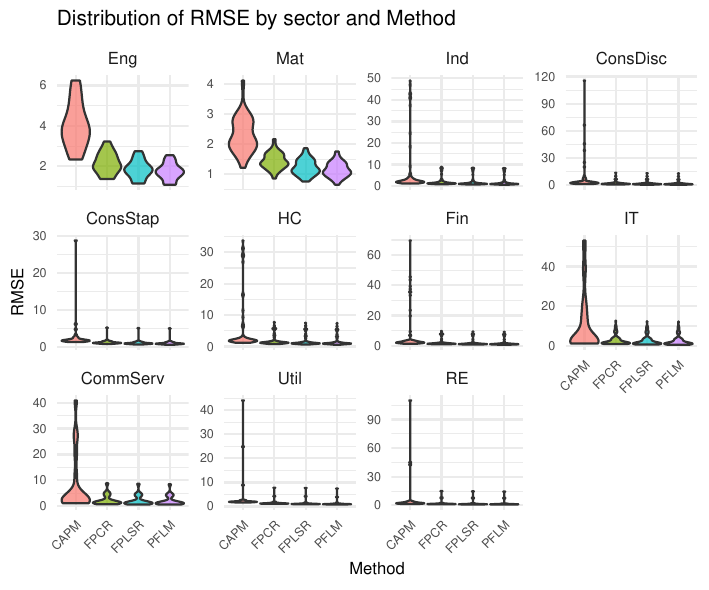}}
\\
\subfloat[RMSPE value]
{\includegraphics[width=8.46cm]{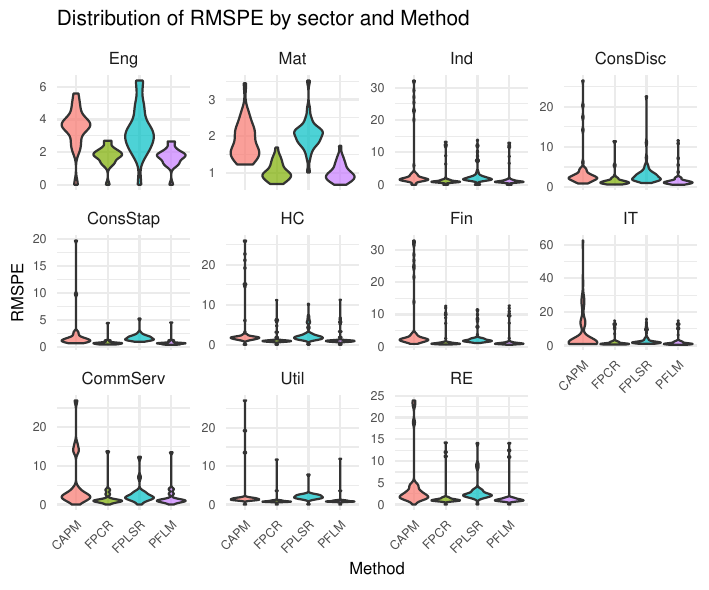}}
\caption{Violin plots of $R^2$, RMSE, and RMSPE values obtained from the classical CAPM and functional CAPM with different estimation methods according to various sectors.}\label{fig:error_2020}
\end{figure}

As a graphical display, we present violin plots in Figure~\ref{fig:error_2020} to compare $R^2$, RMSE, and RMSPE values obtained from the CAPM and functional CAPM with different estimation methods for various industry sectors.

In summary, the 2020 data results align closely with those from 2021, highlighting the consistency and robustness of the functional methods. The PFLM method excels in in-sample performance, while the FPCR method proves to be the most effective for out-of-sample forecasting. These functional methods consistently outperform the traditional CAPM by leveraging high-frequency intraday data, capturing richer information, and mitigating measurement errors. Sensitivity analysis supports the validity of functional CAPM approaches in varying market conditions and confirms their applicability to a wide range of industries.

\newpage
\section{Functional principal component regression}\label{sec:app_a}

Functional principal component analysis extracts latent components based on the largest variance explained in each predictor and response variable \citep[see, e.g.,][]{Shang14, WCM16}. Let us denote 
\begin{align*}
\text{C}_{\bm{\Y}^{j,c}}(v_1, v_2) &= \text{cov}[\bm{\Y}^{j,c}(v_1), \bm{\Y}^{j,c}(v_2)] \\
\mathcal{C}_{\bm{\X}^{c}}(u_1, u_2)&= \text{cov}[\bm{\X}^{c}(u_1), \bm{\X}^{c}(u_2)] 
\end{align*}
as the empirical covariance functions of $\bm{\Y}^{j,c}(v)$ and $\bm{\X}^{c}(u)$, respectively. By Mercer's theorem \citep{Mercer09}, we have the following representations for the covariance functions:
\begin{align*}
\mathcal{C}_{\bm{\Y}^{j,c}}(v_1, v_2) &= \sum_{k \geq 1} \widehat{\lambda}_k^j \widehat{\phi}_k^j(v_1) \widehat{\phi}_k^j(v_2), \\
\mathcal{C}_{\bm{\X}^{c}}(u_1, u_2) &= \sum_{m \geq 1} \widehat{\delta}_m \widehat{\psi}_m(u_1) \widehat{\psi}_m(u_2),
\end{align*}
where $\{\widehat{\phi}_k^j(v): k=1, 2, \ldots\}$ and $\{\widehat{\psi}_m(u): m=1, 2, \ldots\}$ are the empirical orthonormal eigenfunctions corresponding to the estimated eigenvalues $\{\widehat{\lambda}_1^j\geq \widehat{\lambda}_2^j\geq \dots \}$ and $\{\widehat{\delta}_1\geq \widehat{\delta}_2\geq \dots\}$. In practice, most of the variability in a functional variable can be captured by the first few eigenfunctions. 

There are at least five approaches for selecting the number of retained principal components:
\begin{inparaenum}[(1)]
\item the scree plot or the fraction of variance explained by the first several functional principal components \citep{Chiou12};
\item the pseudo-versions of the Akaike information criterion and Bayesian information criterion \citep{YMW05};
\item the cross-validation with one-curve-leave-out \citep{RS91};
\item the bootstrap technique \citep{HV06}; and
\item the eigenvalue ratio criterion \citep{LRS20}.
\end{inparaenum}
In this study, we chose the first $K_j$ and $M$ eigenfunctions, which explain at least 95\% of the total variation in the data, to project the functional response and functional predictor onto orthonormal basis expansions \citep{BSR2023}. 

By Karhunen-Lo\`{e}ve expansion \citep{Karhunen47, Loeve78}, the realizations of the functional variables can be approximated by:
\begin{align}
\Y_t^{j,c}(v) &\approx \sum_{k=1}^{K_j} \widehat{c}^j_{t,k} \widehat{\phi}^j_k(v) = \bm{\widehat{C}}^j_t \bm{\widehat{\Phi}}^j(v), \label{eq:fun_Y} \\
\X_t^{c}(u) &\approx \sum_{m=1}^{M} \widehat{d}_{t,m} \widehat{\psi}_m(u) = \bm{\widehat{D}}_t \bm{\widehat{\Psi}}(u), \label{eq:fun_X} 
\end{align}
where $\widehat{c}^j_{t,k} = \int_{\mathcal{I}} \Y_t^{j,c}(v) \widehat{\phi}_k^j(v) dv$ and $\widehat{d}_{t,m} = \int_{\mathcal{I}} \X_t^{c}(u) \widehat{\psi}_m(u) du$ are the projections of $\Y_t^{j,c}(v)$ and $\X_t^{c}(u)$ onto orthonormal basis functions, respectively. Let $\bm{\widehat{C}}^j_t=(\widehat{c}_{t,1}^j, \widehat{c}_{t,2}^j, \dots, \widehat{c}_{t,K_j}^j)$ and $\bm{\widehat{D}}_t=(\widehat{d}_{t,1}, \widehat{d}_{t,2}, \dots, \widehat{d}_{t,M})$ be two matrices of estimated principal component scores, and let $\bm{\widehat{\Phi}}^j(v) = [\widehat{\phi}_1^j(v), \widehat{\phi}_2^j(v), \dots,\widehat{\phi}_{K_j}^j(v)]^{\top}$ and $\bm{\widehat{\Psi}}(u) = [\widehat{\psi}_1(u), \widehat{\psi}_2(u), \dots,\widehat{\psi}_M(u)]^{\top}$ be two matrices of empirical orthonormal basis functions (e.g., functional principal components). In addition, the error function $\varepsilon^j_t(v)$ admits the expansion with the same basis function in $\Y^{j,c}_t(v)$ as follows:
\begin{equation*}
\varepsilon^j_t(v) \approx \sum_{k=1}^{K_{j}} \widehat{e}_{t,k} \widehat{\phi}_k^j(v) = \bm{\widehat{e}}_t^j \bm{\widehat{\Phi}}^j(v),
\end{equation*}
where the random error function $\widehat{e}_{t,k} = \int_{\mathcal{I}} \widehat{\varepsilon}_t^j(v) \widehat{\phi}_k^j(v) dv$ is i.i.d., and denote $\bm{\widehat{e}}_t = (\widehat{e}_{t,1}, \widehat{e}_{t,2}, \dots, \widehat{e}_{t,K_j})$ and $\bm{\widehat{\Phi}}^j(v) = [\widehat{\phi}_1^j(v), \widehat{\phi}_2^j(v), \dots,\widehat{\phi}_{K_j}^j(v)]^{\top}$.

The bivariate coefficient function can be represented by:
\begin{align}
\widehat{\beta}^j(u,v) &= \sum_{m=1}^{M} \sum_{k=1}^{K_{j}} \widehat{\beta}_{m,k} \widehat{\psi}_m(u) \widehat{\phi}^j_k(v) \notag\\
&= \bm{\widehat{\Psi}}^\top(u) \bm{\widehat{\beta}}^j \bm{\widehat{\Phi}}^j(v),\label{eq:fun_beta}
\end{align}
where $\widehat{\beta}_{m,k} = \int_{\mathcal{I}} \int_{\mathcal{I}} \widehat{\beta}^j(u,v) \widehat{\psi}_m(u) \widehat{\phi}_k^j(v) du dv$ and $\bm{\widehat{\beta}}^j$ is a $(M\times K_j)$ real-valued matrix. 

By substituting the basis expansion forms of the functional variables in~\eqref{eq:fun_Y},~\eqref{eq:fun_X} and~\eqref{eq:fun_beta} into the functional CAPM, we have the following representation:
\begin{align}
\bm{\widehat{C}}_t^j \bm{\widehat{\Phi}}^j(v) &= \int_{\mathcal{I}} \bm{\widehat{D}}_t \bm{\widehat{\Psi}}(u) \bm{\widehat{\Psi}}^\top(u) \bm{\widehat{\beta}}^j \bm{\widehat{\Phi}}^j(v)du + \bm{\widehat{e}}_t \bm{\widehat{\Phi}}^j(v),\notag\\
&= \bm{\widehat{D}}_t \bm{\widehat{\beta}}^j \bm{\widehat{\Phi}}^j(v) \int_{\mathcal{I}}\bm{\widehat{\Psi}}(u) \bm{\widehat{\Psi}}^\top(u) du + \bm{\widehat{e}}_t^j \bm{\widehat{\Phi}}^j(v).\label{eq:4.1}
\end{align} 
Due to the orthonormality property of bases $\bm{\widehat{\Phi}}^j(v)$ and $\bm{\widehat{\Psi}}(u)$, we have $= \int_{\mathcal{I}} \bm{\widehat{\Psi}}(u) \bm{\widehat{\Psi}}^\top(u) du = ~\bm{1}$. By dividing $\widehat{\bm{\Phi}}^j(v)$ from both sides,~\eqref{eq:4.1} reduces to
\begin{equation*}
\bm{\widehat{C}}_t^j = \bm{\widehat{D}}_t \bm{\widehat{\beta}}^j+\bm{\widehat{e}}_t^j,\qquad t=1,2,\dots,n.
\end{equation*}
Let $\widehat{\bm{D}}=(\widehat{\bm{D}}_1, \widehat{\bm{D}}_2, \dots,\widehat{\bm{D}}_n)$ and $\widehat{\bm{C}}^j = (\widehat{\bm{C}}^j_1, \widehat{\bm{C}}^j_2, \dots,\widehat{\bm{C}}^j_n)$, $\widehat{\bm{\beta}}^j$ can be estimated via ordinary least squares
\begin{equation*}
\widehat{\bm{\beta}}^j=(\widehat{\bm{D}}^{\top}\widehat{\bm{D}})^{-1}\widehat{\bm{D}}^{\top}\widehat{\bm{C}}^j.
\end{equation*}
The estimate of the bivariate coefficient function is obtained as
\begin{equation}
\widehat{\beta}^j(u,v) = \bm{\widehat{\Psi}}^\top(u) \widehat{\bm{\beta}}^j \bm{\widehat{\Phi}}^j(v).\label{eq:beta_FPCR}
\end{equation}

\newpage
\section{Functional partial least squares regression}\label{sec:app_b}

From~\eqref{eq:B_spline_1} and~\eqref{eq:B_spline_2}, the functional CAPM can also be expressed as:
\begin{equation}
\bm{\widehat{Z}}_t^j\bm{\widehat{\Pi}}^j(v) = \int_{\mathcal{I}} \bm{\widehat{A}}_t\bm{\widehat{\Gamma}}(u)\widehat{\beta}^j(u,v)du+e_t^j(v) \label{eq:addition}
\end{equation}

We multiple~\eqref{eq:addition} by $\widehat{\bm{\Pi}}^j(v)$ on both sides, and integrating with respect to $v$, we obtain  
\begin{equation*}
\int_{\mathcal{I}} \bm{\widehat{Z}}_t^j\bm{\widehat{\Pi}}^j(v)\bm{\widehat{\Pi}}^j(v)dv=\int_{\mathcal{I}}\int_{\mathcal{I}} \bm{\widehat{A}}_t\bm{\widehat{\Gamma}}(u)\widehat{\beta}^j(u,v)\bm{\widehat{\Pi}}^j(v)dudv + \int_{\mathcal{I}} e_t^j(v)\bm{\widehat{\Pi}}^j(v) dv 
\end{equation*}
Denote by $\widehat{\bm{\Pi}}^j = \int_{\mathcal{I}} \bm{\widehat{\Pi}}^j(v) [\bm{\widehat{\Pi}}^j(v)]^{\top} dv$ and $\widehat{\bm{\Gamma}} = \int_{\mathcal{I}} \bm{\widehat{\Gamma}}(u) \bm{\widehat{\Gamma}}^\top(u) du$ the symmetric matrices of the inner products of the $B$-spline basis functions. Let $\epsilon^{j}_{t} = \int_{\mathcal{I}} e_{t}^{j}(v)\bm{\widehat{\Pi}}^{j}(v) dv$. From~\eqref{eq:beta_FPCR}, we observe $\widehat{\beta}^{j}(u,v) = \bm{\widehat{\Gamma}}^\top(u) \widehat{\bm{\beta}}^j \bm{\widehat{\Pi}}^{j}(v)$, then
\begin{align}
\bm{\widehat{Z}}_{t}^{j}\widehat{\bm{\Pi}}^{j}&=\widehat{\bm{A}}_t\widehat{\bm{\beta}}^{j}\widehat{\bm{\Pi}}^{j}\widehat{\bm{\Gamma}}+\epsilon_{t}^{j} \notag\\
\bm{\widehat{Z}}_{t}^{j}(\widehat{\bm{\Pi}}^{\frac{1}{2}})^{j}[(\widehat{\bm{\Pi}}^{\frac{1}{2}})^{j}]^{\top}&= \bm{\widehat{A}}_t\widehat{\bm{\beta}}^{j}(\widehat{\bm{\Pi}}^{\frac{1}{2}})^j[(\widehat{\bm{\Pi}}^{\frac{1}{2}})^j]^{\top}(\widehat{\bm{\Gamma}}^{\frac{1}{2}})[(\widehat{\bm{\Gamma}}^{\frac{1}{2}})]^{\top}+\epsilon^{j}_{t}. \label{eq:B-splines}
\end{align}
By multiplying both sides of~\eqref{eq:B-splines} by $[(\bm{\Pi}^{-\frac{1}{2}})^{j}]^{\top}$, we obtain
\begin{align*}
\bm{\widehat{Z}}_t^j(\widehat{\bm{\Pi}}^{\frac{1}{2}})^j&=\bm{\widehat{A}}_t\widehat{\bm{\beta}}^j(\widehat{\bm{\Pi}}^{\frac{1}{2}})^j(\widehat{\bm{\Gamma}}^{\frac{1}{2}})(\widehat{\bm{\Gamma}}^{\frac{1}{2}})^{\top}+\epsilon_t^j\times [(\bm{\Pi}^{-\frac{1}{2}})^j]^{\top} \\
&=\bm{\widehat{A}}_t\widehat{\bm{\Gamma}}^{\frac{1}{2}}\widehat{\bm{\beta}}^j(\widehat{\bm{\Pi}}^{\frac{1}{2}})^j(\widehat{\bm{\Gamma}}^{\frac{1}{2}})^{\top}+\tilde{\epsilon}_t^j.
\end{align*}

Denote by $\bm{\widehat{Z}}^j=(\widehat{Z}_1^j, \widehat{Z}_2^j, \dots, \widehat{Z}_n^j)$, $\bm{\widehat{A}}=(\widehat{A}_1, \widehat{A}_2, \dots, \widehat{A}_n)$ and $\bm{\tilde{\epsilon}}^j = (\tilde{\epsilon}_1^j, \tilde{\epsilon}_2^j,\dots,\tilde{\epsilon}_n^j)$. Using a multivariate partial least squares, we obtain
\begin{equation}\label{eq:PLS_basis}
(\bm{\widehat{Z}}^j)^{\top} (\widehat{\bm{\Pi}}^{\frac{1}{2}})^j = \bm{\widehat{A}}^{\top} \widehat{\bm{\Gamma}}^{\frac{1}{2}} \bm{\Omega}^j + \bm{\tilde{\epsilon}^j},
\end{equation}
where 
\begin{equation}
\bm{\Omega}^j=\widehat{\bm{\beta}}^j(\widehat{\bm{\Pi}}^{\frac{1}{2}})^j(\widehat{\bm{\Gamma}}^{\frac{1}{2}})^{\top}, \label{eq:omega_fun}
\end{equation}
and $\bm{\tilde{\epsilon}^j}$ represent the coefficient and residual matrices, respectively. From~\eqref{eq:PLS_basis}, we apply the multivariate partial least-squares method of \cite{BS20} to estimate the coefficient matrix $\bm{\Omega}^j$. From~\eqref{eq:omega_fun}, we obtain the estimated $\widehat{\bm{\beta}}^j$, 
\begin{equation}
\widehat{\bm{\beta}}^j = \widehat{\bm{\Gamma}}^{-\frac{1}{2}}\bm{\Omega}^j(\widehat{\bm{\Pi}}^{-\frac{1}{2}})^j. \label{eq:11}
\end{equation}

The bivariate regression coefficient function $\beta(u,v)$ in the functional CAPM can be estimated from~\eqref{eq:beta_FPCR} and~\eqref{eq:11} as follows:
\begin{equation*}
\widehat{\beta}^j(u,v) = \bm{\widehat{\Gamma}}^{\top}(u) \left[\widehat{\bm{\Gamma}}^{-\frac{1}{2}}\bm{\Omega}^j(\widehat{\bm{\Pi}}^{-\frac{1}{2}})^j\right] \bm{\widehat{\Pi}}^j(v).
\end{equation*}

\newpage
\section{Penalized function-on-function regression}\label{sec:app_c}

A penalized function-on-function (PFLM) regression methodology is adopted to estimate the bivariate regression coefficient function $\beta^j(u,v)$. In this approach, the regularized estimate of $\beta^j(u,v)$ can be obtained by minimizing the objective function:
\begin{equation*}
\underset{\begin{subarray}{c}
  \beta^j(u,b) 
  \end{subarray}}{\arg\min}~ \sum_{j} \left[ \Y^{j,c}_t(v) - \int_{\mathcal{I}} \beta^j(u,v) \X^c_t(u) du \right]^2 + \frac{\kappa}{2} \mathcal{J}(\beta),
\end{equation*}
where $\mathcal{J}$ represents a roughness penalty on $\beta^{j}(u,v)$, and $\kappa$ serves as the smoothing parameter controlling the degree of shrinkage in $\beta^j(u,v)$.

As outlined in Section~\ref{sec:B-spline}, we make the assumption that $\Y^{j,c}_t(v)$ and $\X^c_t (u)$ are densely observed, such that $\Y^{j,c}_t(v_i) = \Y^{j,c}_t(v_j)$ and $\X^c_t (u_i) = \X^c_t (u_j)$ for $i = 1, \ldots, 78$. The regression coefficient function is assumed to be represented by tensor product B-spline basis functions $\lbrace \widehat{\pi}^{j}_{k}(v), k = 1, \ldots, 20 \rbrace$ and $\lbrace \widehat{\gamma}_{m}(u), m = 1, \ldots, 20 \rbrace$, given by:
\begin{equation*}
\beta^j(u,v) = \sum_{k=1}^{20} \sum_{m=1}^{20} \widehat{b}_{km} \widehat{\pi}^j_k(v) \widehat{\gamma}_m(u),
\end{equation*}
where $\widehat{b}_{km}$ represents the B-spline basis expansion coefficients. Let $\Delta_{r}$ denote the length of the $r$\textsuperscript{th} interval in $\mathcal{I}$, such that $\Delta_r = i_{r+1} - i_r$. Following a similar approach to \cite{ivanescu2015}, we employ numerical integration to approximate $\int_{\mathcal{I}} \beta^{j}(u,s) \X^{c}_{t}(u) du$ as follows:
\begin{align}
\int_{\mathcal{I}} \beta^j(u,v) \X^c_t(u) du & \approx \sum_{i=1}^{19} \Delta_r \beta^j(u_i,v) \X^c_t(u_i) \nonumber \\
& = \sum_{r=1}^{19} \Delta_r \sum_{k=1}^{20} \sum_{m=1}^{20} \widehat{b}_{km} \widehat{\pi}^j_k(v) \widehat{\gamma}_m(u_i) X^c_t(u_i) \nonumber \\
& = \sum_{k=1}^{20} \sum_{m=1}^{20} \widehat{b}_{km} \widehat{\pi}^j_k(v) \widetilde{\widehat{\gamma}}_{m} \label{eq:numint},
\end{align}
where $\widetilde{\widehat{\gamma}}_{m} = \sum_{r=1}^{19} \Delta_r \widehat{\gamma}_m(u_i) X^c_t(u_i)$. Then, the functional CAPM model is approximated as follows:
\begin{equation*}
\Y^{j,c}_t(v) = \sum_{k=1}^{20} \sum_{m=1}^{20} \widehat{b}_{km} \widehat{\pi}^j_k(v) \widetilde{\widehat{\gamma}}_{m} + \varepsilon_t^{j,c}(v).
\end{equation*}

Consider a $20 \times 20$ dimensional matrix of basis expansion coefficients denoted by $\widehat{\bm{b}} = (\widehat{b}_{km})_{km}$. The penalty functional $\mathcal{J}(\beta)$ is approximated in the following manner:
\begin{align}\label{eq:p2}
\widetilde{\mathcal{J}}(\beta) &= \int_{\mathcal{I}} \int_{\mathcal{I}} \left[ \frac{\partial^2}{\partial v^2} \beta(u,v) \right]^2 du dv + \int_{\mathcal{I}} \int_{\mathcal{I}} \left[ \frac{\partial^2}{\partial u^2} \beta(u,v) \right]^2 du dv \notag\\
&= \widehat{\bm{b}}^\top (\widehat{\bm{\Gamma}} \otimes \bm{P}_y + \bm{P}_x \otimes \widehat{\bm{\Pi}}^j) \widehat{\bm{b}},
\end{align}
where $\widehat{\bm{\Pi}}^j = \int_{\mathcal{I}} \bm{\widehat{\Pi}}^j(v) (\bm{\widehat{\Pi}}^j)^\top(v) dv$, $\widehat{\bm{\Gamma}} = \int_{\mathcal{I}} \bm{\widehat{\Gamma}}(u) (\bm{\widehat{\Gamma}})^\top(u) du$, and $\bm{P}_y$ and $\bm{P}_x$ are the penalty matrices, with $(k k^{\prime})\textsuperscript{th}$ and $(m m^{\prime})\textsuperscript{th}$ entries; $P_{y, k k^{\prime}} = \int_{\mathcal{I}} (\widehat{\pi}_k^j)^{(2)}(v) (\widehat{\pi}_{k^{\prime}}^j)^{(2)}(v) dv$ and $P_{x, m m^{\prime}} = \int_{\mathcal{I}} \widehat{\gamma}_m^{(2)}(u) \widehat{\gamma}_{m^{\prime}}^{(2)}(u) du$ for $k, k^{\prime}, m, m^{\prime} = 1, \ldots, 20$.

By employing the estimated integral in~\eqref{eq:numint} and the approximated penalty functional described in equation \eqref{eq:p2}, the estimation of $\widehat{\bm{b}}$ can be achieved by minimizing:
\begin{equation}\label{eq:obj2}
\underset{\begin{subarray}{c}
  \widehat{\bm{b}}
  \end{subarray}}{\arg\min}~ \sum_{j} \sum_{i=1}^{78} \left[ \Y^{j,c}_t(v_i) - (\widetilde{\widehat{\bm{\gamma}}}^\top \otimes (\bm{\widehat{\Pi}}^j)^\top (v_i)) \widehat{\bm{b}} \right]^2 + \frac{\kappa}{2} \widetilde{\mathcal{J}}(\beta),
\end{equation}
where $\widetilde{\widehat{\bm{\gamma}}} = [\widetilde{\widehat{\gamma}}_{1}, \ldots, \widetilde{\widehat{\gamma}}_{20}]^T$. Accordingly, the regularized estimate of $\beta^j(u,v)$ is obtained as follows:
\begin{equation*}
\widehat{\beta}^{j}(u,v) = (\widehat{\bm{\Gamma}}(u) \otimes (\bm{\widehat{\Pi}}^{j})^{\top} (v)) \widehat{\bm{b}}^{*},
\end{equation*}
where $\widehat{\bm{b}}^*$ is the estimates of $\widehat{\bm{b}}$ obtained by minimizing~\eqref{eq:obj2}.

The estimation of $\widehat{\bm{b}}$ is achieved through a penalized least squares approach. In this method, determining the optimal value of $\kappa$ is crucial for efficient estimation results. Various information criteria, including the Bayesian information criterion (BIC), generalized cross-validation, and modified Akaike information criterion, are available for this purpose. We propose using BIC to identify the optimal smoothing parameter due to its simplicity and computational efficiency. The BIC for determining the optimal smoothing parameter can be computed as follows:
\begin{equation*}
\text{BIC}(\kappa) = N \times \ln \bigg \Vert \sum_{j} \left[ \Y^{j,c}_t(v) - \widehat{\Y}^{j,c}_{t, \kappa}(v) \right] \bigg \Vert_2^2 + \ln(N),
\end{equation*}
where $\widehat{\Y}^{j,c}_{t, \kappa}(v)$ represents the estimate of $\Y^{j,c}_t(v)$ with the smoothing parameter $\kappa$. It's important to note that the optimal value of the penalty parameter $\kappa$ is determined using a standard grid-search approach with a predefined set of candidate values for $\kappa$.

\newpage
\bibliographystyle{agsm}

\end{document}